\documentclass[conference]{IEEEtran}
\IEEEoverridecommandlockouts
\usepackage{amsmath,amssymb,amsthm}
\usepackage{booktabs,graphicx,xspace,xcolor}
\usepackage{enumitem}
\setlist[itemize]{leftmargin=*}
\setlist[enumerate]{leftmargin=*}
\usepackage{algorithm,algpseudocode}
\algrenewcommand{\algorithmicrequire}{\textbf{Input:}}
\algrenewcommand{\algorithmicensure}{\textbf{Output:}}
\usepackage[hidelinks]{hyperref}
\newtheorem{theorem}{Theorem}
\newtheorem{lemma}{Lemma}
\newtheoremstyle{reviewcorollary}
  {\topsep}{\topsep}{\itshape}{}{\bfseries}{\textcolor{black}{.}}{5pt plus 1pt minus 1pt}
  {\textcolor{black}{\thmname{#1}\thmnumber{ #2}}\thmnote{ {\normalfont(#3)}}}
\theoremstyle{reviewcorollary}
\newtheorem{corollary}{Corollary}
\theoremstyle{definition}
\newtheorem{definition}{Definition}
\newtheorem{problem}{Problem}
\theoremstyle{plain}
\definecolor{writingpurple}{RGB}{0,0,0}

\newcommand{\method}{\texttt{SIG}\xspace}
\newcommand{\methodfullname}{\underline{S}ource-\underline{I}ntegrated \underline{G}reedy}
\newcommand{\sand}{\texttt{SandIMIN}\xspace}
\newcommand{\sandminus}{\texttt{SandIMIN-}\xspace}

\title{Efficient Influence Minimization via Source-Integrated Sampling}
\author{\IEEEauthorblockN{Jingbang Chen\textsuperscript{*}}
\IEEEauthorblockA{\textit{CUHK-Shenzhen \& SLAI}\\
chenjb@cuhk.edu.cn}
\and
\IEEEauthorblockN{Junyao Wang}
\IEEEauthorblockA{\textit{HK PolyU}\\
junya.wang@connect.polyu.hk}
\thanks{*Corresponding author. This work has been submitted to the IEEE for possible publication. Copyright may be transferred without notice, after which this version may no longer be accessible.}}
\begin{document}
\maketitle
\bstctlcite{icdebstcontrol}
\begin{abstract}
\looseness=-1
Influence minimization by node blocking is a fundamental
problem in graph data management that seeks a small set of vertices
whose removal minimizes the expected spread from a fixed source set. It supports decisions about where to intervene in large
interaction networks to control harmful propagation, such as the spread
of rumors and infectious diseases.
Under the Independent Cascade (IC) model, selecting blockers requires
comparing a large number of candidate sets, and estimating their effects on
propagation can be costly. We propose \method (\methodfullname),
a sampling framework for efficiently finding high-quality
approximate solutions. {\color{black}Our key idea is to group the possible sets of neighbors
directly activated by the sources according to their effects on blocker
selection.
\method represents these groups in a compact weighted tree and computes
their probabilities directly, allowing each sample to account for all
source outcomes without enumerating them. We prove that \textcolor{black}{\method produces unbiased estimates
of the lower-bound objective} without increasing variance for fixed
blocking plans, and establish a high-probability quality guarantee
through independent validation.} Extensive experiments on real-world networks show that \method
has the lowest mean total running time in all main comparisons, with
speedups of up to $583\times$ over the fastest baseline.
\textcolor{black}{\method produces solutions competitive with those of prior methods and
achieves the highest influence reduction in multiple settings,
including on many of the largest networks, where several baselines exceed
the runtime limit.} Additional experiments show that \method maintains consistently
low running times as graph size and the blocking budget increase\textcolor{black}{,
while delivering stronger blocking solutions at larger budgets.}
\end{abstract}
\section{Introduction}
\label{sec:introduction}

\suppressfloats[t]
\begin{figure}[t]
\centering
\includegraphics[width=\columnwidth]{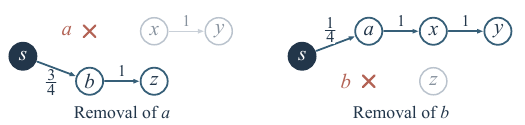}
\caption{\textcolor{black}{Expected spread under two blocking choices ($k=1$).}}
\label{fig:intro-blocking}
\end{figure}

{\color{black}
Influence minimization by node blocking is a well-studied problem in
graph data management~\cite{shi2019adaptive,xie2023vertex,sandimin2024,chang2025timin,wang2025tcim}.
It supports propagation-control decisions such as account suspension
to limit misinformation~\cite{mccabe2024deplatforming}, account blocking
to slow the spread of event-related posts, and road-checkpoint placement in epidemic
simulations~\cite{teng2023iminimize}.
Data scientists often study this problem under the
\textit{Independent Cascade (IC)} model~\cite{kempe2003influence,sandimin2024},
in which a newly active vertex makes one independent attempt along
each outgoing edge to activate an inactive neighbor.
Each attempt succeeds with the edge's transmission probability, which
can be estimated from activity logs such as records of users joining
online communities~\cite{goyal2010probabilities}.
Given a fixed source set and a budget $k$, the goal is to choose up to
$k$ other vertices whose removal before propagation minimizes the
expected number of vertices reached from the source set.
Blocking a vertex removes it and its incident edges. \textcolor{black}{The expected spread after blocking depends on how likely each
remaining vertex is to be reached from the sources~\cite{jin2011distance,ke2019reliability}.}
\textcolor{black}{In Figure~\ref{fig:intro-blocking}, the expected spread outside $s$ is
$3(1/4)+2(3/4)=2.25$ before blocking. For $k=1$, removing $a$ leaves
$2(3/4)=1.50$, whereas removing $b$ leaves $3(1/4)=0.75$.
Thus, $b$ is the better choice despite its shorter branch.}

A well-chosen blocking plan can substantially reduce propagation
with only a small intervention budget.
On a social network of 1.13 million YouTube users,
blocking 100 users, fewer than $0.01\%$ of the network, lowers the
estimated number reached outside the source set from about $80{,}900$
to $45{,}100$, a reduction of $44.3\%$.
Despite its importance, finding an effective plan is difficult and
costly: estimating the effects of many possible plans requires substantial computation,
especially on large graphs.
}

{\color{black}
Choosing which vertices to block is difficult because deletions can
work jointly: a target may remain reachable after either of two
deletions but become disconnected after both.
Existing algorithms reuse sampled propagation graphs to score
deletions~\cite{xie2023vertex}, or use sampled lower and upper bounds
to guide selection and assess solution
quality~\cite{sandimin2024,wang2025tcim}.
\textcolor{black}{The expected number of targets protected by at least one chosen
blocker acting alone is a lower bound on influence reduction.}
\textcolor{black}{For a sampled target, source-edge outcomes can change which
vertices individually protect it, even when all other edge outcomes are fixed.
Many samples may therefore be needed to compare candidate sets reliably,
increasing the cost of selection.}

\textcolor{black}{To reduce the number of samples needed for these comparisons, we represent
source-edge outcomes compactly within each sample.}
\textcolor{black}{Outcomes that activate the sampled target are grouped by which
vertices could stop that activation on their own.}
\textcolor{black}{Our key observation is that these vertex sets form paths in a
single tree.}
\textcolor{black}{A compact weighted tree stores these groups and their probabilities,
allowing one sample to score many blocking plans without enumerating
the \textcolor{black}{exponentially many} source outcomes.}

As a result, we develop a new framework named \method
(\methodfullname). \textcolor{black}{It builds a blocking plan one vertex at a time, choosing the vertex
with the largest additional sampled score. Independent samples then
evaluate the plan's influence reduction and bound the best achievable
influence reduction.}
We evaluate \method on real-world networks to assess its efficiency,
solution quality, and scalability.

Our contributions are summarized as follows:
\begin{itemize}
\item \textcolor{black}{We develop \method, a sampling framework for influence minimization
under the IC model. It groups source outcomes by which vertices can
individually protect a target. It maintains these groups and their
probabilities in compact weighted trees for repeated comparisons of
blocking plans.}
\item \textcolor{black}{We prove that \method's sampling mechanism produces unbiased estimates
of the lower-bound objective, with no greater variance for fixed
blocking plans than estimates based on sampled source outcomes.
We also establish high-probability bounds on the selected plan's true
influence reduction and the optimum, thereby proving the validity of
its approximation certificates.}
\item \textcolor{black}{Extensive experiments\footnote{\textcolor{black}{Code and experiment scripts are available at \url{https://github.com/chenjb1997/sig-artifact}.}} on real-world networks show that
\method has the lowest mean total running time in all main comparisons,
with speedups of up to $583\times$ over the \textcolor{black}{fastest baseline}.}
\textcolor{black}{\method produces solutions competitive with those of prior methods and
achieves the highest influence reduction in multiple settings,
including on many of the largest networks, where several baselines exceed
the runtime limit.}
\item \textcolor{black}{Additional experiments show that \method maintains consistently low
running times as graph size and the blocking budget increase, \textcolor{black}{and
delivers stronger blocking solutions with larger budgets.}}
\textcolor{black}{We also study parameter sensitivity and \textcolor{black}{assess the effectiveness of
the sampling and validation designs through an ablation study.}}
\end{itemize}

\noindent\textbf{Outline.} The rest of the paper is organized as
follows. We review related work in Section~\ref{sec:related}.
Section~\ref{sec:preliminaries} introduces the necessary notation
and the influence minimization problem.
Section~\ref{sec:method-source-integrated} presents \method, and
Section~\ref{sec:analysis-source-integrated} gives its theoretical
analysis. Section~\ref{sec:evaluation} reports the empirical
evaluations. We conclude in Section~\ref{sec:conclusion}.
}

\section{Related Work}
\label{sec:related}

\textcolor{black}{In this section, we review related work on influence
minimization and intervention, influence models and constraints, and
influence maximization and estimation.}

\paragraph{Influence Minimization and Intervention}
Xie et al. introduce \texttt{AdvancedGreedy} and \texttt{GreedyReplace}
for IC vertex blocking~\cite{xie2023vertex}, and study node and edge
blocking under IC and LT~\cite{xie2025blocking}.
Their dominator trees evaluate single-vertex gains on sampled live graphs.
\sand bounds the non-submodular IC blocking objective by submodular
lower and upper functions~\cite{sandimin2024}.
\texttt{CBFM} develops sampled bounds for time-critical
blocking~\cite{wang2025tcim}; with an infinite deadline and unit
activation values, its lower bound becomes CP coverage and its upper
bound uses a selected live path.
Minimum-cardinality formulations seek \textcolor{black}{vertex sets} satisfying a prescribed
influence reduction~\cite{pham2019blocking,sandimin2026}.
Observation-based policies adapt blocking to the observed
diffusion~\cite{shi2019adaptive}, while time-aware temporary blocking
is studied under the Temporal Linear Threshold
model~\cite{chang2025timin}.
Other interventions change node features to reduce
diffusion~\cite{xiang2025hypermin}, or add links to increase spread
from fixed seeds~\cite{chen2025link}.
Our setting selects \textcolor{black}{vertices for permanent blocking} before IC propagation begins;
its adaptive stages determine the sampling effort used to select and
validate this set.

\paragraph{Influence Models and Constraints}
\textcolor{black}{Influence optimization has also been studied under
different diffusion models and application constraints.}
Extensions consider topics and message
content~\cite{chen2015topic,li2019multifaceted}, geodemographic
reach~\cite{zhang2020geodemographic}, and coordination-game
propagation~\cite{zhang2017game}.
Other formulations account for seed costs and coverage
requirements~\cite{bian2020budgeted,feng2024seedselection}, capacity
limits~\cite{zhang2023capacity}, fairness~\cite{feng2023fairness}, or
matroid constraints~\cite{huang2024matroid}.
Marketing formulations consider multiple interacting
products~\cite{tsaras2021collective}, seed countering~\cite{xie2024countering},
revenue~\cite{khan2016revenue}, and stakeholder
pricing~\cite{huang2023stakeholders}.
Influence optimization also considers evolving
networks~\cite{ohsaka2016dynamic,zhao2019tracking,huang2019temporal},
closed social networks~\cite{huang2022closed}, and repeated
campaigns~\cite{sun2018multiround}.
Feedback supports adaptive seed
selection~\cite{han2018adaptive,guo2023budgetadaptive,huang2020profit}
and online parameter learning~\cite{wu2019factorization}; robust
formulations account for uncertain diffusion
models~\cite{chen2016robust,he2016robust}.
\textcolor{black}{Our study adopts the commonly used IC
model~\cite{kempe2003influence,sandimin2024}.}

\paragraph{Influence Maximization and Estimation}
\textcolor{black}{Kempe et al. formulate influence maximization under IC and LT and
establish greedy approximation guarantees~\cite{kempe2003influence}.
Reverse influence sampling casts seed selection as coverage of
sampled reverse-reachable sets~\cite{borgs2014rr}.
TIM and IMM develop scalable algorithms with sample-count estimation
and approximation guarantees~\cite{tang2014tim,tang2015imm}.
Later work refines adaptive sampling and stopping
rules~\cite{nguyen2016ssa,huang2017revisiting} and accelerates
reverse-reachable set generation~\cite{guo2020rr}.
Chen et al. analytically average over independent seed activations
within such samples for lattice influence
maximization~\cite{chen2020lattice}.
For node blocking, \method groups source-edge outcomes by common-path
sets and stores the group probabilities on reverse dominator trees
for repeated comparisons of candidate blocker sets.}

\begingroup
\section{Preliminaries}
\label{sec:preliminaries}

\subsection{Notation}
\label{sec:prelim-notation}

We consider a directed graph $G=(V,E)$ with $n=|V|$ vertices and
$m=|E|$ edges. Each edge $e$ has a propagation probability
$p_e\in[0,1]$. The source set $S\subseteq V$ is fixed, and
$C=V\setminus S$ is the set of vertices that can be blocked.
For a graph $H$, $\operatorname{Reach}_H(S)$ denotes the vertices
reachable from $S$, including the sources. We write $H-B$ for the graph
obtained by deleting $B\subseteq C$ and its incident edges.

\begin{definition}[Independent Cascade (IC)~\cite{kempe2003influence}]
\label{def:prelim-ic}
All sources are initially active. For each outgoing edge $e=(u,v)$
to an inactive vertex $v$, a newly active $u$ makes one activation
attempt, independently with probability $p_e$. Active vertices stay
active, and propagation ends when no new vertex is activated.
\end{definition}

Equivalently, a \emph{live-edge graph} $\Phi$ includes each edge $e$
independently with probability $p_e$. Its active vertices are exactly
$\operatorname{Reach}_\Phi(S)$~\cite{kempe2003influence}.
The \emph{influence spread} of $S$ is
$\sigma_G(S)=\operatorname{E}_\Phi[|\operatorname{Reach}_\Phi(S)|]$.

\begin{definition}[Common-Path (CP) Set~\cite{sandimin2024}]
\label{def:prelim-cp}
Fix a live-edge graph $\Phi$ and a target $t\in C$ reachable from $S$.
The common-path set $\mathrm{CP}_{\Phi}(t)$ contains the vertices
outside $S$ that lie on every directed path from a source to $t$ in
$\Phi$. It includes $t$. If $t$ is unreachable from $S$, its CP set
is empty.
\end{definition}
A vertex belongs to this set exactly when blocking that vertex alone
prevents an otherwise active $t$ from being activated.
For target $t$ in Figure~\ref{fig:prelim-blocking}(a), when both
edges leaving $s$ are live, the paths $s\to a\to t$ and
$s\to b\to t$ have only $t$ in common outside $S$, so
$\mathrm{CP}_{\Phi}(t)=\{t\}$.
When only $(s,a)$ is live, the CP set is $\{a,t\}$.
This explains the lower bound used in~\cite{sandimin2024}.
A blocker set that intersects the CP set saves $t$ because one of
its vertices already suffices. With both source edges live,
blocking $\{a,b\}$ also saves $t$ by cutting the two paths jointly.

\subsection{Problem Specification}
\label{sec:prelim-problem}

Following the node-blocking formulation
in~\cite{xie2023vertex,sandimin2024}, we select vertices before
propagation starts. Blocking deletes those vertices and their incident
edges; the sources and the probabilities of all other edges stay fixed.

\begin{definition}[Influence Reduction~\cite{sandimin2024}]
\label{def:prelim-reduction}
For a blocker set $B\subseteq C$, the expected number of vertices
saved from activation is
\begin{equation}
\begin{aligned}
 D(B)&=\sigma_G(S)-\sigma_{G-B}(S)\\
 &=\operatorname{E}_\Phi\!\left[
 \left|\operatorname{Reach}_\Phi(S)
 \setminus\operatorname{Reach}_{\Phi-B}(S)\right|\right].
\end{aligned}
\label{eq:method-objective}
\end{equation}
The two reachability sets use the same live-edge graph.
A blocked vertex contributes to $D(B)$ when it would have been active
in the original graph.
\end{definition}

\begin{problem}[Influence Minimization via Node Blocking~\cite{sandimin2024}]
\label{prob:prelim-blocking}
Given $G$, edge probabilities $p$, sources $S$, and an integer budget
$0\leq k\leq |C|$, find
\begin{equation}
 B_D^*\in\operatorname*{arg\,max}_{B\subseteq C,\,|B|\leq k}D(B).
 \label{eq:prelim-problem}
\end{equation}
We denote its value by $\mathrm{OPT}_D=D(B_D^*)$.
\end{problem}

\begin{figure}[t]
\centering
\includegraphics[width=0.94\columnwidth]{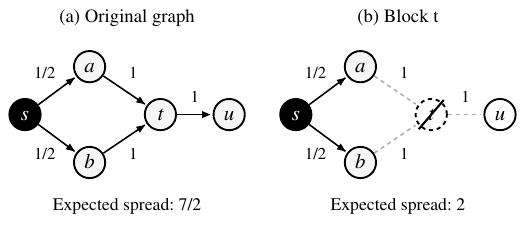}
\caption{Node blocking with source $S=\{s\}$.
Edge labels give propagation probabilities. Blocking $t$ removes
the dashed edges in (b) and leaves $u$ unreachable from $s$.}
\label{fig:prelim-blocking}
\end{figure}

In Figure~\ref{fig:prelim-blocking}(a), $a$ and $b$ are each activated
with probability $1/2$. Both $t$ and $u$ are activated with probability
$1-(1-1/2)^2=3/4$, so $\sigma_G(S)=7/2$.
For $k=1$, blocking $t$ is optimal: only $s,a,b$ can be active,
giving $\sigma_{G-\{t\}}(S)=2$ and $D(\{t\})=3/2$.
Blocking $a$ or $b$ saves $1$ in expectation, while blocking $u$
saves $3/4$. With budget two, blocking $\{a,b\}$ saves all four
non-source vertices whenever they would have been active, giving
$D(\{a,b\})=5/2$.

\subsection{Coverage and Variance}
\label{sec:prelim-math}

A set function $f:2^C\to\mathbb R_{\geq0}$ is \emph{monotone} if
$f(A)\leq f(B)$ for $A\subseteq B$, and \emph{submodular} if
\begin{equation}
 f(A\cup\{v\})-f(A)\geq f(B\cup\{v\})-f(B)
 \label{eq:prelim-submodularity}
\end{equation}
for $A\subseteq B\subseteq C$ and $v\notin B$~\cite{nemhauser1978}.
Influence reduction is monotone but generally not
submodular~\cite{xie2023vertex,sandimin2024}. For example,
$D(\{b\})-D(\varnothing)=1$ in Figure~\ref{fig:prelim-blocking},
whereas $D(\{a,b\})-D(\{a\})=3/2$.

\begin{samepage}
\begin{lemma}[Weighted Coverage And Greedy Maximization~\cite{nemhauser1978}]
\label{l:prelim-coverage}
Let each $v\in C$ cover a fixed finite set $Q_v$ of elements with nonnegative
weights $w_x$. Then
$f(B)=\sum_{x\in\bigcup_{v\in B}Q_v}w_x$ is monotone and submodular,
with $f(\varnothing)=0$. For $1\leq k\leq |C|$, greedily adding a vertex
of maximum marginal gain for $k$ steps gives a set $B_g$ satisfying
\begin{equation}
 f(B_g)\geq\bigl[1-(1-1/k)^k\bigr]
              \max_{B\subseteq C,\,|B|\leq k} f(B).
 \label{eq:prelim-greedy}
\end{equation}
\end{lemma}
\end{samepage}

We use weighted coverage to represent sampled bounds on $D$
in Section~\ref{sec:method-source-integrated}. The lemma gives the
greedy guarantee for these sampled coverage functions.

\begin{lemma}[Variance Reduction by Conditioning~\cite{blackwell1947}]
\label{l:prelim-conditioning}
For a square-integrable random variable $Y$ and a random element $Z$,
let $\bar Y=\operatorname{E}[Y\mid Z]$. Then
$\operatorname{E}\bar Y=\operatorname{E}Y$ and
\begin{equation}
 \operatorname{Var}(Y)=\operatorname{Var}(\bar Y)
       +\operatorname{E}[\operatorname{Var}(Y\mid Z)].
 \label{eq:prelim-total-variance}
\end{equation}
In particular, $\operatorname{Var}(\bar Y)\leq\operatorname{Var}(Y)$.
\end{lemma}

Section~\ref{sec:method-variance} applies this result by conditioning
on a target and a live-edge graph on $C$, then averaging over the
random edges from $S$ to $C$.
\endgroup

\begingroup
\providecommand{\Cref}[1]{Section~\ref{#1}}
\color{black}
\section{Source-Integrated Sampling Algorithm}
\label{sec:method-source-integrated}
\label{sec:source-integrated-algorithm}

We formally propose our main algorithm, \method (\methodfullname), for
Problem~\ref{prob:prelim-blocking}. We now give an overview of its
framework and key idea.

\subsection{Overview}
\label{sec:sig-overview}

\begin{algorithm}[t]
\caption{Source-integrated influence minimization (\method)}
\label{alg:sig}
\small
\begin{algorithmic}[1]
\Require Directed graph $G$, probabilities $p$, sources $S$, budget $k$,
\Statex \hspace{\algorithmicindent} target \textcolor{black}{$\rho\in(0,1]$}, initial sample size $P_0$, stage limit $J$
\Ensure A feasible blocker set with its solution status, or failure
\State Check the input and set $C\gets V\setminus S$ \label{alg:line:prepare}
\State Find the endpoints $F\subseteq C$ of positive-probability
\Statex \hspace{\algorithmicindent} edges from $S$
\State \textbf{if} $|F|\leq k$ \textbf{then return} $(F,\mathrm{optimal})$ \label{alg:line:trivial}
\State \textcolor{black}{Compute source-entry probabilities $q$} \label{alg:line:support}
\State Find $R^*\subseteq C$ reachable from $S$ along edges with $p_e>0$
\State Set $M\gets|R^*|$
\State \textbf{if} $k=0$ or $M=0$ \textbf{then return} $(\varnothing,\mathrm{optimal})$ \label{alg:line:zero}
\State $(B^*,L^*,U^*)\gets(\varnothing,0,M)$
\For{$j=0,\ldots,J-1$}
  \State $P_j\gets 2^j P_0$. Start an empty collection $\mathcal W_j$
  \State Allocate $T_j^{\rm tr}$ targets in $R^*$ as in Section~\ref{sec:method-gates} \label{alg:line:train}
  \For{each allocated target $t$} \Comment{Algorithm~\ref{alg:lower-sketch}}
    \State $W\gets{}$\textcolor{black}{\Call{BuildWeightedSample}{$t,G[C],q$}} \label{alg:line:build}
    \State \textbf{if} $W$ is nonempty \textbf{then} add $W$ to $\mathcal W_j$
  \EndFor
  \State Choose $k$ blockers $B_j$ by greedy coverage on $\mathcal W_j$ \label{alg:line:choose}
  \State Freeze $B_j$ and release $\mathcal W_j$
  \State Draw $P_j$ fresh iid validation samples \label{alg:line:fresh}
  \State Compute $(L_j,U_j)$ for $D(B_j)$ and $\mathrm{OPT}_D$ \label{alg:line:validate}
  \State $(B',L')\gets(B^*,L^*)$
  \State \textbf{if} $j=0$ or $L_j>L^*$ \textbf{then} $(B',L')\gets(B_j,L_j)$ \label{alg:line:combine}
  \State $U'\gets\min\{U^*,U_j\}$ \label{alg:line:upper}
  \State \textbf{if} $L'>U'$ \textbf{then report} validation failure \textbf{and halt} \label{alg:line:failure}
  \State $(B^*,L^*,U^*)\gets(B',L',U')$ \label{alg:line:commit}
  \If{\textcolor{black}{$L^*\geq\rho U^*$}} \label{alg:line:stop}
    \State \Return $(B^*,\mathrm{target\ attained})$
  \EndIf
\EndFor
\State \Return $(B^*,\mathrm{target\ unmet})$ \label{alg:line:cap}
\end{algorithmic}
\end{algorithm}

\textcolor{black}{Algorithm~\ref{alg:sig} seeks a prescribed fraction $\rho\in(0,1]$
of the optimal influence reduction. It repeats the following three
steps with increasing sample sizes.}
\begin{enumerate}
 \item \textbf{Build weighted samples:}
 Sample live edges outside $S$ and summarize CP coverage in weighted trees
 (Section~\ref{sec:method-gates}).
 \item \textbf{Select blockers:}
 Repeatedly select the vertex covering the largest remaining weight,
 up to $k$ blockers
 (Section~\ref{sec:method-stopping}).
 \item \textbf{Validate and stop:}
 Use fresh samples to bound candidate quality and the optimum.
 Stop when these bounds certify the target ratio; otherwise,
 increase the sample size (Section~\ref{sec:method-witness}).
\end{enumerate}

The main improvement lies in sampling: we average over source-edge
outcomes within each sample. Fix a target and the live edges among
vertices outside $S$. The source edges have exponentially many possible
outcomes, but their nonempty CP sets are paths in a single tree. There
are at most as many such paths as vertices that can reach the target.
We assign each path its probability and store these weights on the
tree. A blocker set covers the paths that it intersects, so one
weighted tree scores many blocker sets. For each fixed set, this
coverage is the conditional mean of its CP coverage over the source-edge
outcomes. Section~\ref{sec:method-gates} gives the construction, and
Section~\ref{sec:analysis-source-integrated} establishes its representation
and variance bounds. For independent validation, we also average
source-edge outcomes in a path-based upper bound.

\subsection{Building Weighted Samples}
\label{sec:method-gates}
\label{sec:method-soft-cp}

This step represents each training draw as a weighted tree whose
subtree weights support greedy blocker selection.

Before sampling, Algorithm~\ref{alg:sig} sets $C=V\setminus S$ and finds
the endpoints $F\subseteq C$ of positive-probability edges from $S$.
If $|F|\leq k$, blocking $F$ cuts every path from the sources into $C$.
Otherwise, it computes source-entry probabilities and finds the set
$R^*\subseteq C$ reachable from $S$ along positive-probability edges,
with $M=|R^*|$. If $k=0$ or $M=0$, the optimal influence reduction is zero.
For each $a\in C$, the probability that at least one edge from $S$ to
$a$ is live is
\begin{equation}
 q_a=1-\prod_{e=(s,a),\,s\in S}(1-p_e).
 \label{eq:method-gates}
\end{equation}
We call $a$ a successful source entry when this event occurs.
These events are mutually independent and independent of the live
edges in $G[C]$ because they involve disjoint edge sets.

For one training draw with target $t\in R^*$,
let $H$ denote its live-edge realization on $C$. We explore only the
reverse region $R=R_H(t)$, which contains $t$ and all vertices that can
reach $t$ in $H$. We write $A$ for the random set of
successful source entries in $R$.
The target is active exactly when $A\ne\varnothing$.
We build the dominator tree $T$ rooted at $t$ in the reverse of $H[R]$.
A vertex $a$ lies in the subtree $T_b$ rooted at $b$ exactly when
every path from $a$ to $t$ passes through $b$.
For nonempty $A$, the CP vertices are the common
ancestors of $A$ in $T$. They form the path from $t$ to the lowest common
ancestor $\operatorname{LCA}_T(A)$ of the entries in $A$, including
both endpoints.

We represent the CP lower bound of \sand~\cite{sandimin2024} by assigning
probabilities to these paths. Conditional on $t,H$, let $\lambda_v$ be
the probability that $A$ is nonempty and $\operatorname{LCA}_T(A)=v$.
A blocker set $B$ intersects this path exactly when
$\operatorname{LCA}_T(A)$ lies in $T_b$ for some $b\in B\cap R$.
The conditional CP coverage of $B$ is therefore
\begin{equation}
 \ell_{t,H}(B)=
 \sum_{v\in\bigcup_{b\in B\cap R}T_b}\lambda_v .
 \label{eq:method-soft-coverage}
\end{equation}
Theorem~\ref{t:method-cp-representation} proves this representation
and its equivalence to the CP lower bound.

The weight computation uses
$\eta_v=1-\prod_{a\in T_v}(1-q_a)$, the probability of at least one
successful entry in $T_v$, and
$O_v=\prod_{a\in R\setminus T_v}(1-q_a)$, the probability that all
entries in $R\setminus T_v$ fail. We denote by $h_v$ the probability
that at least two child subtrees of $v$ contain a successful entry.
The entries in $A$ have lowest common ancestor
$v$ exactly when all entries in $R\setminus T_v$ fail and either $v$ is
a successful source entry or \textcolor{black}{its source-entry event fails}
and at least two child subtrees
of $v$ contain a successful entry.
Thus
\begin{equation}
 \lambda_v=O_v[q_v+(1-q_v)h_v].
 \label{eq:method-atoms}
\end{equation}

\begin{algorithm}[t]
\caption{\textcolor{black}{\textsc{BuildWeightedSample}$(t,G[C],q)$}}
\label{alg:lower-sketch}
\small
\begin{algorithmic}[1]
\Require Target $t$, graph $G[C]$ with edge probabilities $p_e$,
\Statex \hspace{\algorithmicindent} \textcolor{black}{source-entry probabilities $q$}
\Ensure A weighted tree represented by its subtree masses, or an empty tree
\State Sample the live reverse region $R$ of $t$ in $G[C]$ \label{sk:draw}
\If{\textcolor{black}{$q_v=0$} for every $v\in R$} \label{sk:zero}
  \State \Return an empty tree
\EndIf
\State Build the reverse dominator tree $T$ rooted at $t$ \label{sk:tree}
\State Compute the singleton probabilities $c_v$ in $T$ \label{sk:single}
\For{each $v$ in postorder of $T$}
   \State Use the child success probabilities to find $\eta_v,h_v$ \label{sk:child}
   \State Set $\lambda_v\gets 0$
   \If{$\eta_v>0$}
      \State $\lambda_v\gets (c_v/\eta_v)[q_v+(1-q_v)h_v]$ \label{sk:atom}
   \EndIf
   \State $m_v\gets\lambda_v+\sum_{u\text{ child of }v}m_u$ \label{sk:mass}
\EndFor
\State \Return \textcolor{black}{$T$ with $m_v$, omitting zero-mass subtrees} \label{sk:return}
\end{algorithmic}
\end{algorithm}

Algorithm~\ref{alg:lower-sketch}, called at line~\ref{alg:line:build} of
Algorithm~\ref{alg:sig}, implements this construction.
The routine samples the reverse region $R$. \textcolor{black}{If all source-entry
probabilities $q_v$ in $R$ are zero, it returns an empty tree.}
Otherwise, it builds $T$ and uses failure products
to compute the singleton protection probabilities $c_v$, which satisfy
$c_v=O_v\eta_v$.
A postorder traversal computes $\eta_v,h_v$ from $q_v$ and the child
subtree success probabilities and evaluates
\eqref{eq:method-atoms} using $O_v=c_v/\eta_v$ when $\eta_v>0$.
Vertices with $\eta_v=0$ receive $\lambda_v=0$.
The routine stores subtree masses
\mbox{$m_v=\lambda_v+\sum_{u\text{ child of }v}m_u$} for greedy coverage.
A draw returning an empty tree still counts toward the training sample size.
\textcolor{black}{We defer the computation details for
lines~\ref{sk:single}, \ref{sk:child}, \ref{sk:atom}, and~\ref{sk:mass}
of Algorithm~\ref{alg:lower-sketch} to \Cref{sec:method-tree-weights},
which also proves the $O(|R|)$ time bound.}

We assemble these trees into a fresh training collection at each stage.
With input sample budget $P_0\geq2$ and stage limit $J\geq1$, we set
\begin{equation}
 P_j=2^jP_0,\qquad j=0,\ldots,J-1.
 \label{eq:adaptive-defaults}
\end{equation}
If $P_j<M$, targets are drawn independently and uniformly from $R^*$.
Otherwise, each target is sampled once per round for as many complete
rounds as fit the budget, giving each target the same number of draws.
Each call to \textsc{BuildWeightedSample} uses fresh live-edge outcomes.
The training count is
\begin{equation}
 T_j^{\mathrm{tr}}=
 \begin{cases}
 P_j & \text{if }P_j<M,\\
 M\lfloor P_j/M\rfloor & \text{if }P_j\geq M.
 \end{cases}
 \label{eq:adaptive-training-count}
\end{equation}

\color{black}
\subsection{Selecting Blockers}
\label{sec:adaptive-schedule}
\label{sec:method-stopping}

This step constructs a candidate blocker set $B_j$
by greedy coverage on the training trees. For $\theta=T_j^{\mathrm{tr}}$
training draws, the sampled lower objective is
\begin{equation}
 \widehat F_j(B)=\frac{M}{\theta}\sum_{i=1}^{\theta}\ell_i(B).
 \label{eq:method-sample-objectives}
\end{equation}
Each allocated draw counts toward $\theta$, including a draw with zero
mass. Under balanced sampling the coefficient is the reciprocal of the
number of complete rounds.

At line~\ref{alg:line:choose}, greedy starts from
the empty set and adds $k$ distinct vertices from $C$ one at a time,
using the same training collection. Each blocker covers a subtree in
every training tree that contains it. At each iteration, the algorithm
adds an unchosen vertex that covers the most additional weight across
the training trees. A priority queue maintains the marginal gains.
Lemma~\ref{l:prelim-coverage} gives the approximation guarantee for
exact greedy selection on the sampled objective
in~\eqref{eq:method-sample-objectives}.
The resulting candidate $B_j$ remains fixed while independent samples
estimate its true influence reduction $D(B_j)$, including protection
that requires multiple blockers to act jointly.

\definecolor{writingpurple}{RGB}{0,0,0}
\subsection{Independent Validation and Stopping}
\label{sec:method-witness}
\label{sec:adaptive-moments}
\label{sec:adaptive-fractional}
\label{sec:adaptive-numeric-endpoints}
\label{sec:adaptive-combination}

This step uses independent samples to lower-bound
$D(B_j)$ and upper-bound $\mathrm{OPT}_D$. A candidate is certified
when its lower bound is at least $\rho$ times an upper bound on
$\mathrm{OPT}_D$.
At lines~\ref{alg:line:fresh}--\ref{alg:line:validate}, we draw
$T=P_j$ independent uniform targets from $R^*$ and a fresh live-edge
graph on $C$ for each target. These draws are independent of training.

For a draw $(t,H)$ and a blocker set $B$, let $R=R_H(t)$ and
$R_B=R_{H-B}(t)$, with $R_B=\varnothing$ if $t\in B$.
The two products below give the probabilities
that $t$ is inactive after and before blocking $B$. Their difference
is the probability that $B$ saves $t$.
\begin{equation}
 d_{t,H}(B)=\prod_{a\in R_B}(1-q_a)-\prod_{a\in R}(1-q_a).
 \label{eq:method-conditional-true}
\end{equation}
Summing $\operatorname{E}_H d_{t,H}(B)$ over all targets gives
$D(B)$ of Definition~\ref{def:prelim-reduction}. Thus the validation
mean for $B_j$, multiplied by $M$, estimates its influence reduction.

To bound $\mathrm{OPT}_D$, we construct a
path-coverage upper bound that applies to every feasible blocker set.
For a validation draw $(t,H)$, each source-edge outcome that
activates $t$ is assigned one live path from a successful entry to $t$.
We define $u_{t,H}(B)$ as the probability that $B$ intersects this
selected path. Saving $t$ requires cutting every live path from its
successful entries, so $u_{t,H}(B)$ upper-bounds $d_{t,H}(B)$.
We construct a shortest-path tree $P$ toward $t$
in $H[R]$. A reverse breadth-first
search computes distances to $t$. Each nonroot vertex chooses the
out-neighbor with the smallest original ID among those one step closer
to $t$. $P_a$ denotes the tree path from a
source entry $a$ to $t$, including both endpoints. We order the source entries
in $R$ as $a_1,\dots,a_g$ by increasing distance and then by original ID.
Each source-edge outcome with a successful entry in $R$ is assigned the path
of its first successful entry in this order. The event that $a_i$ succeeds
and all preceding entries fail has probability
\begin{equation}
 \omega_{a_i}=q_{a_i}\prod_{j<i}(1-q_{a_j}).
 \label{eq:method-first-gate}
\end{equation}
Thus $P_{a_i}$ is selected with probability
$\omega_{a_i}$. Summing the weights of paths intersected by $B$ gives
\begin{equation}
 u_{t,H}(B)=\sum_i\omega_{a_i}\,
              \mathbf{1}\{B\cap P_{a_i}\ne\varnothing\}.
 \label{eq:method-witness}
\end{equation}
In $P$, blocking a vertex $b$ covers the weights
of entries in the subtree rooted at $b$.
\texttt{CBFM} constructs its selected-path upper bound from sampled
propagation outcomes~\cite{wang2025tcim}. Here, the path weights average
over the source-edge outcomes.

The path weights sum to the target's conditional
activation probability,
$\pi_{t,H}=1-\prod_{a\in R}(1-q_a)$.
This construction satisfies
\begin{equation}
 \ell_{t,H}(B)\leq d_{t,H}(B)\leq u_{t,H}(B)\leq\pi_{t,H}
 \label{eq:method-sandwich}
\end{equation}
for every blocker set on the same draw.
Theorem~\ref{t:method-witness-sandwich} establishes these inequalities.

The validation batch gives the values
$d_i(B_j)$ for the candidate. Let $B_D^*$ denote a fixed optimal set
of the original instance, with ties resolved before sampling.
The weighted paths give upper bounds $Q_1$ on its sample sum and
$Q_2$ on its sum of squares. The latter yields the variance bound needed for
empirical Bernstein validation.

For each observation, $u_i$ is the path coverage in~\eqref{eq:method-witness},
and $\pi_i$ is the target's conditional
activation probability. Since $0\leq d_i(B)\leq u_i(B)\leq\pi_i$,
\begin{equation}
 d_i(B)^2\leq \pi_i u_i(B).
 \label{eq:moment-pointwise}
\end{equation}
Scaling the weights on the path tree by $\pi_i$ therefore gives a
coverage upper for squared protection. Define the two coverage sums
$f_1(B)=\sum_i u_i(B)$ and $f_2(B)=\sum_i\pi_i u_i(B)$, where both
sums run over the validation batch.
We seek computable upper bounds satisfying
\begin{equation}
 Q_h\ \geq\ \max_{B\subseteq C,\,|B|\leq k} f_h(B),
 \qquad h\in\{1,2\}.
 \label{eq:moment-global}
\end{equation}

For $X_i=d_i(B_D^*)$, feasibility of $B_D^*$
gives $\sum_iX_i\leq f_1(B_D^*)\leq Q_1$ and
$\sum_iX_i^2\leq f_2(B_D^*)\leq Q_2$.
The empirical mean of these observations is therefore at most
$\min\{1,Q_1/T\}$. For $T\geq2$, their sample variance satisfies
\begin{equation}
 s_*^2=\frac{\sum_i X_i^2-T\overline X^2}{T-1}
 \leq v_*=\min\!\left\{\frac{Q_2}{T-1},\frac{T}{4(T-1)}\right\}.
 \label{eq:moment-opt-variance}
\end{equation}
The first bound follows by dropping the nonnegative
centering term. The range bound follows from
$X_i^2\leq X_i$ and $\overline X(1-\overline X)\leq1/4$.
For the frozen candidate $B_j$, we can compute its own sample variance
directly from the sum and sum of squares of $d_i(B_j)$. We denote
\textcolor{black}{an upper bound on this variance by $v_j$}, applying the same
range bound.

We compute $Q_1$ and $Q_2$ by bounding the two
coverage sums separately. For either $f=f_h$, we construct
a computable upper bound using the candidate as an anchor, denoted
$A=B_j$ in the following bounds. For each
$v\in C$, let $s_v=f(\{v\})$ be its singleton weight and
$r_v=f(A\cup\{v\})-f(A)$ its remaining weight after covering $A$.
In particular $r_v=0$ for $v\in A$. Let $\mathrm{Top}_k(z)$ denote
the sum of the $k$ largest entries of a vector $z$ over $C$.
By submodularity, for every feasible $B$,
\begin{equation}
 f(B)\leq\sum_{v\in B}s_v,\qquad
 f(B)\leq f(A)+\sum_{v\in B}r_v .
 \label{eq:method-sampled-upper}
\end{equation}
Monotonicity gives $f(B)\leq f(A\cup B)$,
and submodularity bounds the added coverage by the marginal gains $r_v$.
These two modular bounds overcount different
parts of the coverage. Mixing their vertex coefficients before taking
the largest $k$ entries can give a smaller upper bound~\cite{iyer2013semidifferential}.

For $0\leq\lambda\leq1$, set
$z_v^\lambda=(1-\lambda)s_v+\lambda r_v$. Lemma~\ref{t:fractional-upper}
gives the mixed upper bound
\begin{equation}
 f(B)\leq Q_\lambda=\lambda f(A)+\mathrm{Top}_k(z^\lambda).
 \label{eq:fractional-upper}
\end{equation}

For each $f_h$, \method takes the smallest of the total path
weight, $T$, and~\eqref{eq:fractional-upper} at
$\lambda\in\{0,1/4,1/2,3/4,1\}$. The total weights are $\sum_i\pi_i$
for $f_1$ and $\sum_i\pi_i^2$ for $f_2$. Each interior mixture scans
the full candidate domain, including members of the anchor, because
their singleton weights still contribute. Since $f_2(B)\leq f_1(B)$,
we further set $Q_2\gets\min\{Q_2,Q_1,T\}$.
Each bound holds for every feasible set on the
batch, so taking the minimum gives $Q_1,Q_2$ satisfying
\eqref{eq:moment-global}.

A blocked reverse traversal gives $d_i(B_j)$, and the
path tree supplies the two coverage sums.

We now use the sample mean and variance bounds to compute the confidence
bounds at line~\ref{alg:line:validate}.
The confidence bounds apply to the true-protection
observations $d_i(B_j)$ for the fixed candidate and $d_i(B_D^*)$ for
the fixed optimal set. Here \textcolor{black}{$x^-\in[0,1]$} is a lower bound on the sample mean
of $d_i(B_j)$, while \textcolor{black}{$y^+=\min\{1,Q_1/T\}$ is an}
upper bound on the sample mean of $d_i(B_D^*)$. The sample variance
bounds are $v_j$ for the candidate and $v_*$
in~\eqref{eq:moment-opt-variance} for the fixed optimal set.
Chernoff and empirical Bernstein each provide
a lower bound for the candidate and an upper bound for the optimum.
Each of the four bounds uses failure probability
\begin{equation}
 \delta=\frac{1}{4nJ}.
 \label{eq:adaptive-spending}
\end{equation}
With four bounds per stage and at most $J$
stages, this allocation makes the completed-stage bounds simultaneously
valid with probability at least $1-1/n$.
Here $J$ is fixed before sampling. \textcolor{black}{The Chernoff bounds use
$c\geq\log(1/\delta)/T$ and the following endpoint functions.
For brevity, we defer their derivation to
\Cref{sec:method-chernoff-bounds}.}
\begin{align}
 b_-(x,c)&=\max\{0,x-\sqrt{2cx}\},\nonumber\\
 b_+(x,c)&=\min\{1,x+c+\sqrt{c^2+2cx}\}.
 \label{eq:adaptive-endpoints}
\end{align}
The empirical Bernstein bound~\cite{maurer2009} uses the radius
\begin{equation}
 r(v)=\sqrt{\frac{2v\log(2/\delta)}{T}}+
       \frac{7\log(2/\delta)}{3(T-1)}.
 \label{eq:moment-eb-radius}
\end{equation}
With $[z]_+=\max\{0,z\}$, \textcolor{black}{the four bounds are}
\begin{equation}
\begin{aligned}
 [L_j^{\rm C},U_j^{\rm C}]&=M[b_-(x^-,c),b_+(y^+,c)],\\
 L_j^{\rm EB}&=M[x^--r(v_j)]_+,\\
 U_j^{\rm EB}&=M\min\{1,y^++r(v_*)\}.
\end{aligned}
\label{eq:adaptive-stage-bounds}
\end{equation}
The upper formulas are nondecreasing in the
sample mean and variance, so substituting $y^+$ and $v_*$ gives
computable upper bounds on $\mathrm{OPT}_D$.

We combine each completed stage's bounds by taking the larger lower
bound $L_j=\max\{L_j^{\rm C},L_j^{\rm EB}\}$ and the smaller upper
bound $U_j=\min\{U_j^{\rm C},U_j^{\rm EB}\}$. Across completed stages
through $j$, lines~\ref{alg:line:combine}--\ref{alg:line:commit} set
\begin{equation}
\begin{aligned}
 i_j&=\min\operatorname*{arg\,max}_{h\le j}L_h,&
 B_j^\star&=B_{i_j},\\
 L_j^\star&=L_{i_j},&
 U_j^\star&=\min_{h\le j} U_h .
\end{aligned}
\label{eq:adaptive-merged}
\end{equation}
Each lower bound $L_h$ is paired with its
candidate $B_h$, whereas all upper bounds $U_h$ apply to the same
$\mathrm{OPT}_D$. With the bounds simultaneously valid,
$D(B_j^\star)\geq L_j^\star$ and $\mathrm{OPT}_D\leq U_j^\star$.
We update the stored candidate and bounds after the
full validation batch and the consistency check at line~\ref{alg:line:failure}.
\textcolor{black}{At line~\ref{alg:line:stop}, target attainment is reported when}
\begin{equation}
 \textcolor{black}{L_j^\star\geq\rho U_j^\star.}
 \label{eq:adaptive-stop}
\end{equation}
At the stage limit, line~\ref{alg:line:cap}
returns the set with the largest completed lower bound and reports
target unmet, meaning that the requested ratio remains uncertified.
\textcolor{black}{An interruption or failed consistency check terminates}
the run with a failure status.
Theorem~\ref{t:adaptive-simultaneous} establishes
the quality guarantee for reported attainment.

\begingroup
\color{black}
\subsection{Computing the Tree Weights}
\label{sec:method-tree-weights}
\label{sec:td-atoms}
We give the computation deferred from \Cref{sec:method-gates}.
In Algorithm~\ref{alg:lower-sketch}, this computes $c_v$ at
line~\ref{sk:single}, $\eta_v,h_v$ at line~\ref{sk:child},
$\lambda_v$ at line~\ref{sk:atom}, and $m_v$ at line~\ref{sk:mass}.
Given the reverse dominator tree $T$ rooted at $t$ and the
source-entry probabilities $q_v$, the following lemmas identify
these quantities and compute them in $O(|R|)$ time.

\begin{lemma}[Tree-Weight Identities]
\label{l:tree-weight-identities}
\textcolor{black}{The weights in~\eqref{eq:method-atoms} are}
nonnegative and satisfy
$m_v=\sum_{x\in T_v}\lambda_x=O_v\eta_v=c_v$ for every $v\in R$.
In particular, $m_t=\pi_{t,H}$.
\end{lemma}
\begin{proof}
The events $\{A\ne\varnothing,\operatorname{LCA}_T(A)=x\}$,
indexed by $x\in R$, partition the target's activation event.
The lowest common ancestor of $A$ lies in $T_v$ exactly when all
source entries outside $T_v$ fail and at least one inside entry
succeeds. Independence gives probability $O_v\eta_v$.
This is also the event that blocking $v$ alone saves $t$, so its
probability is $c_v$. At the root, $O_t=1$ and $\eta_t=\pi_{t,H}$.
\end{proof}

\begin{lemma}[Linear-Time Weight Computation]
\label{l:tree-weight-time}
Given $T$ and $q_v$, all quantities
$\eta_v,h_v,c_v,\lambda_v,m_v$ can be computed in $O(|R|)$
time and space under unit-cost arithmetic.
\end{lemma}
\begin{proof}
\textcolor{black}{Here $\operatorname{ch}(v)$ denotes the children of $v$,
and $F_v=1-\eta_v$.}
A bottom-up traversal uses
$F_v=(1-q_v)\prod_{u\in\operatorname{ch}(v)}F_u$.
A top-down traversal starts with $O_t=1$ and uses
$O_u=O_v(1-q_v)\prod_{w\in\operatorname{ch}(v)\setminus\{u\}}F_w$
for each child $u$.
Prefix and suffix products compute the sibling products in
constant time per child, even when some factors are zero.

To compute $h_v$, initialize $(p_0,p_1,p_2)=(1,0,0)$ for zero,
one, or at least two successful child subtrees.
For a child $u$ with $h=\eta_u$, update simultaneously
$p'_0=p_0(1-h)$, $p'_1=p_1(1-h)+p_0h$, and
$p'_2=p_2+p_1h$.
After all children, $h_v=p_2$.
Set $c_v=m_v=O_v\eta_v$ and evaluate
\eqref{eq:method-atoms} for $\lambda_v$.
Every tree edge is processed a constant number of times,
establishing the time and space bounds.
\end{proof}
\endgroup

\begingroup
\color{black}
\subsection{Deriving the Chernoff Bounds}
\label{sec:method-chernoff-bounds}
\label{sec:td-chernoff}
We derive the endpoint functions $b_-(x,c)$ and $b_+(x,c)$
used in \Cref{sec:method-witness}.
For a validation batch of size $T$, let $X_i=d_i(B)\in[0,1]$
be the independent true-protection observations of a fixed
blocker set $B$. Write $\overline X=T^{-1}\sum_iX_i$ and
$\mu=\operatorname{E}\overline X=D(B)/M$.
Let $x\in[0,1]$ denote a possible value of the sample mean.
Convexity gives $e^{zX_i}\leq1+X_i(e^z-1)$ for every real $z$.
Independence and $1+y\leq e^y$ then yield
\begin{equation}
 \operatorname{E}e^{z\sum_iX_i}\leq \exp\{T\mu(e^z-1)\}.
 \label{eq:td-mgf}
\end{equation}
For $x,\mu>0$, applying Markov's inequality with $z=\log(x/\mu)$
bounds the corresponding tail by $\exp[-T\mu \psi(x/\mu)]$, where
$\psi(y)=y\log y-y+1$. Since $\psi(1)=\psi'(1)=0$ and $\psi''(y)=1/y$,
integration between $1$ and $y$ gives
\[
 \psi(y)\geq (y-1)^2/(2\max\{1,y\}).
\]
For $x>\mu$ and $x<\mu$, respectively, this yields
\begin{align}
 \Pr(\overline X\geq x)&\leq\exp[-T(x-\mu)^2/(2x)],\nonumber\\
 \Pr(\overline X\leq x)&\leq\exp[-T(\mu-x)^2/(2\mu)].
 \label{eq:adaptive-tail-relaxations}
\end{align}
Let $c\geq\log(1/\delta)/T$. Inverting the upper tail gives
the lower endpoint $\max\{0,x-\sqrt{2cx}\}$ for $\mu$.
For the lower tail, the boundary equation
$(\mu-x)^2=2c\mu$ has upper root
$x+c+\sqrt{c^2+2cx}$.
Clipping this root at $1$ gives the upper endpoint.
\textcolor{black}{These are the two functions
in~\eqref{eq:adaptive-endpoints}, each with one-sided failure
probability at most $\delta$.}
\textcolor{black}{The formulas extend continuously to $x=0$. If $\mu=0$, all
observations are zero almost surely.}
Both functions are nondecreasing in $x$, which justifies
substituting $x^-$ for the candidate and $y^+$ for the fixed
optimum in~\eqref{eq:adaptive-stage-bounds}.

\endgroup

\endgroup

\begingroup
\definecolor{writingpurple}{RGB}{0,0,0}
\color{black}
\section{\textcolor{black}{Analysis of Source-Integrated Sampling}}
\label{sec:analysis-source-integrated}
\label{sec:adaptive-source-integrated}

We establish the coverage representation, variance reduction, and
solution quality of Algorithm~\ref{alg:sig}, then analyze its time
and space requirements.

\subsection{\textcolor{black}{Coverage Representation}}
\label{sec:analysis-coverage}

The lower construction groups source-edge outcomes by their common-path
sets. The following result shows that its tree weights represent the
entire distribution for a fixed target and sampled graph.

\begin{theorem}[Conditional Common-Path Representation]
\label{t:method-cp-representation}
Fix a target $t\in C$ and a live-edge graph $H$ on $C$. Let $A$ be the
successful source entries in $R_H(t)$ and let $T$ be the reverse
dominator tree rooted at $t$. For $v\in R_H(t)$, let $\lambda_v$
be the probability that $A$ is nonempty and its lowest common
ancestor in $T$ is $v$. These weights are nonnegative and sum to
$\pi_{t,H}$. Equation~\eqref{eq:method-soft-coverage} holds
simultaneously for every $B\subseteq C$.
\textcolor{black}{The sum
$\sum_{t\in C}\operatorname{E}_H\ell_{t,H}(B)$ equals the CP lower
bound used by \sand~\cite{sandimin2024}.}
\end{theorem}
\begin{proof}
For nonempty $A$, a vertex $b$ belongs to the common-path set exactly
when every path from any vertex in $A$ to $t$ passes through $b$.
In $T$, these vertices are the common ancestors of $A$. They form
the path from $t$ to $Z=\operatorname{LCA}_T(A)$.
This path intersects $B$ exactly when $Z$ belongs to a subtree rooted
at a vertex of $B$. The possible locations of $Z$ partition the
event $A\ne\varnothing$, whose probability is $\pi_{t,H}$.
Adding their probabilities proves~\eqref{eq:method-soft-coverage}.
When $A=\varnothing$, the common-path set is empty. For each target,
a common-path hit means that some selected vertex alone saves it.
\textcolor{black}{Taking expectations and summing over targets
gives this lower bound.}
\end{proof}

The upper construction chooses one live path for each possible first
successful source entry. Its relation to true protection holds for
all blocker sets on the same sampled graph.

\begin{theorem}[Bounds on Target Protection]
\label{t:method-witness-sandwich}
\textcolor{black}{For fixed $t,H$ and every $B\subseteq C$,}
$\ell_{t,H}(B)\leq d_{t,H}(B)\leq u_{t,H}(B)\leq\pi_{t,H}$.
Moreover, $u_{t,H}(B)\leq u^{\rm LRR}_{t,H}(B)$, where $u^{\rm LRR}$
is the conditional hit probability of \sand's local reverse-reachable
set~\cite{sandimin2024}. \textcolor{black}{In the same realization,
the LRR set consists of the active vertices that can reach $t$ along
live paths within $C$.}
\end{theorem}
\begin{proof}
\textcolor{black}{For each source-edge realization, a common-path
hit means that some vertex of $B$ alone saves $t$, so blocking $B$
also saves it.}
If $B$ saves $t$, it intersects every live path from a successful
source entry to $t$, including the path selected by the upper
construction. Each vertex on this selected path is active, belongs
to $C$, and reaches $t$ through active vertices in $C$. The path
is therefore contained in the local reverse-reachable set used
by \sand~\cite{sandimin2024}. Each hit event requires the target
to be active before blocking. Taking conditional probabilities
gives all the inequalities in~\eqref{eq:method-sandwich}.
\end{proof}

\begingroup
\color{black}
In addition, we can prove the following exactness result for
directed out-forests.
\begin{corollary}[Exactness on Directed Out-Forests]
\label{l:method-forest}
\label{t:method-forest}
If the graph induced by $C$ is a directed out-forest after removing
self-loops and identifying parallel edges, then
$\ell_{t,H}(B)=d_{t,H}(B)=u_{t,H}(B)$ for every $t,H,B$.
\end{corollary}
\endgroup

The next lemma justifies the mixed upper bound used for the two
validation coverage sums.

\begin{lemma}[Mixed Coverage Bound]
\label{t:fractional-upper}
For any nonnegative weighted coverage function $f$, anchor $A$, and
$\lambda\in[0,1]$, the value $Q_\lambda$
in~\eqref{eq:fractional-upper} satisfies $f(B)\leq Q_\lambda$
for every $B\subseteq C$ with $|B|\leq k$.
\end{lemma}
\begin{proof}
Let $s_v=f(\{v\})$ and $r_v=f(A\cup\{v\})-f(A)$ as in the
algorithm. \textcolor{black}{The singleton sum counts every
element covered by $B$ at least once, so
$f(B)\leq\sum_{v\in B}s_v$.}
Monotonicity and submodularity give
$f(B)\leq f(A\cup B)\leq f(A)+\sum_{v\in B}r_v$.
\textcolor{black}{Their weighted combination gives
\[
 f(B)\leq\lambda f(A)+
 \sum_{v\in B}\bigl[(1-\lambda)s_v+\lambda r_v\bigr].
\]}
The resulting sum over $B$ is at most the sum of the $k$
largest coefficients because every coefficient is nonnegative.
This is~\eqref{eq:fractional-upper}.
\end{proof}

Every candidate bound is valid on the realized batch for every
feasible set. Thus the algorithm can use its sampled candidate as
the anchor and take the minimum over the prescribed grid.

\subsection{\textcolor{black}{Variance Reduction}}
\label{sec:analysis-variance}
\label{sec:method-variance}

Source integration replaces a random common-path hit by its conditional
probability. The reduction in variance also applies to comparisons
between fixed blocker sets.

\begin{theorem}[Variance Reduction for Fixed Comparisons]
\label{t:method-variance}
For fixed blocker sets $B_1,\ldots,B_j$, let $I$ be their vector of
common-path-hit indicators on one draw of $(t,H,A)$, and let
$L=\operatorname{E}[I\mid t,H]$. Then
\textcolor{black}{$\operatorname{Cov}(I)-\operatorname{Cov}(L)$
is positive semidefinite.}
\end{theorem}
\begin{proof}
\textcolor{black}{With $Y=c^{\mathsf T}I$ and $z=c^{\mathsf T}L$
for a fixed $c\in\mathbb R^j$, we have
$z=\operatorname{E}[Y\mid t,H]$.
Lemma~\ref{l:prelim-conditioning} gives
$\operatorname{Var}(Y)-\operatorname{Var}(z)
 =\operatorname{E}[\operatorname{Var}(Y\mid t,H)]\geq0$.
Since this holds for every $c$, the covariance difference is
positive semidefinite.}
\end{proof}

For two sets, the theorem bounds the variance of their score difference.
The same conditioning argument applies to selected-path hits with
conditional means $u_{t,H}(B)$.

\subsection{\textcolor{black}{Solution Quality}}
\label{sec:analysis-quality}

Independent validation gives a confidence bound on the selected set
and an upper bound on the true optimum. The algorithm uses their
ratio to decide when to stop.

\begin{theorem}[Solution Quality]
\label{t:adaptive-simultaneous}
\textcolor{black}{Assume the IC sampling model and fresh independent validation draws
conditional on all preceding work.}
For one invocation, with probability at least $1-1/n$, every completed
sampling stage satisfies
\begin{equation}
 L_j^\star\leq D(B_j^\star)\leq\mathrm{OPT}_D\leq U_j^\star .
 \label{eq:adaptive-simultaneous}
\end{equation}
Consequently, the probability that \method reports attainment for a set
with $D(B_j^\star)<\rho\,\mathrm{OPT}_D$ is at most $1/n$.
The preprocessing returns an optimal set whenever one of its return
conditions holds.
\end{theorem}
\begin{proof}
Condition on the history up to the start of stage $j$'s validation.
Its selected $B_j$ is fixed, and its fresh observations have mean
$D(B_j)/M$. Each lower confidence bound fails with probability at
most $\delta=1/(4nJ)$. Fix an optimal set $B_D^*$ of the original
instance before sampling. Its observations have mean
$\mathrm{OPT}_D/M$. On every batch, their sample mean is at most
$y^+$ and their sample variance is at most $v_*$. Substituting these
bounds into the increasing upper confidence formulas gives failure
probability at most $\delta$ for each upper bound.

Define all potential stage batches in advance. A union bound over the
four bounds at each of the $J$ stages gives total failure probability
at most $4J\delta=1/n$. This covers the completed stages of the
algorithm, including when it stops early. On the simultaneous event,
\eqref{eq:adaptive-merged} pairs the largest lower bound with the set
that earned it and chooses a valid upper bound on $\mathrm{OPT}_D$.
The test in~\eqref{eq:adaptive-stop} then implies
\textcolor{black}{$D(B_j^\star)\geq L_j^\star\geq\rho U_j^\star
\geq\rho\,\mathrm{OPT}_D$}.

For a structural return, deleting the complete set $F$ of vertices
that a source could activate directly saves every vertex in $C$
that could be activated.
If $k=0$ or $M=0$, the optimal influence reduction is zero.
These are the preprocessing return conditions, proving optimality.
\end{proof}

The returned set remains associated with its own lower bound throughout
stage merging. At the stage cap, the same confidence interval describes
its quality.

\subsection{\textcolor{black}{Complexity}}
\label{sec:analysis-complexity}
\label{sec:adaptive-cost}

\begingroup
\color{black}
We analyze the adjacency-list implementation under unit-cost word
arithmetic. For sample $i$, let $a_i$ count examined incoming edges,
$e_i$ stored live edges, and $r_i$ reached vertices, with
$w_i=a_i+e_i+r_i$. The dominator-tree construction cost for a training
sample is $\mathcal D_i$.
At stage $j$, $S_j$ is the total number of stored tree vertices.
$K_j$ counts ancestor updates, incidence-entry visits during score
refreshes, and successor-pointer traversals, including repeated visits.
$H_j$ counts heap operations. With $\mathcal T_j$ and $\mathcal V_j$
indexing the training and validation draws, define the stage work
\begin{equation}
\begin{aligned}
 \Gamma_j={}&n+S_j+K_j+H_j\log(n+1)\\
 &+|C|\log(k+1)+\sum_{i\in\mathcal T_j}(w_i+\mathcal D_i)\\
 &+\sum_{i\in\mathcal V_j}\bigl(w_i+r_i\log(r_i+1)\bigr).
\end{aligned}
\label{eq:sig-stage-cost}
\end{equation}

\begin{theorem}[Time And Space Complexity]
\label{t:sig-cost}
The preprocessing and the first $h$ completed stages, for
$1\leq h\leq J$, take
$O(n+m+\sum_{j<h}\Gamma_j)$ time and
$O(n+m+\max_{j<h}S_j+Jk)$ peak space.
Training and validation each use at most $P_0(2^h-1)$ observations.
A structural return takes $O(n+m)$ time and space.
\end{theorem}
\begin{proof}
Preprocessing takes $O(n+m)$ time and space. The vertex index is
initialized once and reset only at visited vertices, so each reverse
sample costs $O(w_i)$. The training pass constructs its dominator tree
in $\mathcal D_i$ time using $O(r_i)$ working arrays and computes subtree
weights in $O(r_i)$ additional time. Stored trees and incidence lists
occupy $O(S_j)$ space. Each stored entry is cleared at most once, and
each selected-vertex incidence is visited at most once. The heap has
at most one entry per unselected vertex. Bottom-up heap construction
takes $O(n)$ time, and its $H_j$ operations take $O(H_j\log(n+1))$.
Ancestor updates, incidence-entry visits, and successor traversals take
$O(K_j)$ time.

Validation processes one sampled graph at a time. Its traversals and
path ordering cost $O(w_i+r_i\log(r_i+1))$ per sample. The constant
number of top-$k$ computations over $C$ takes
$O(|C|\log(k+1))$ time, with $O(n)$ time and space for score arrays.
Training sketches are released before validation. Together with the
current graph and the $O(Jk)$ stored blocker sets and top-$k$ lists,
this gives the peak-space bound. Summing the stage costs gives the
time bound. The stage budgets sum to
$\sum_{j<h}P_j=P_0(2^h-1)$, and the training count
in~\eqref{eq:adaptive-training-count} is at most $P_j$.
A structural return needs only preprocessing.
\end{proof}
\endgroup
\endgroup

\begingroup
\section{Evaluation}
\label{sec:evaluation}

\begingroup
\providecommand{\Cref}[1]{Section~\ref{#1}}
We evaluate \method's running time and influence reduction on real-world
graphs. We describe the datasets, baselines, and settings
(\Cref{sec:eval-setting}), compare running times at two blocker budgets
(\Cref{sec:eval-main}), and assess the quality of the returned blocker sets
(\Cref{sec:eval-quality}). We then study how running time changes with
graph size (\Cref{sec:eval-scalability}) and blocker budget
(\Cref{sec:eval-budget-scalability}) and how parameter choices affect
performance (\Cref{sec:eval-parameters}).
\textcolor{black}{We close with an ablation study (\Cref{sec:eval-ablation}).}
\par
\endgroup

\subsection{Experiment Setting}
\label{sec:eval-setting}

\paragraph{Datasets}
We use seven SNAP networks~\cite{snapnets}%
\footnote{\url{https://snap.stanford.edu/data/}} and
Flixster~\cite{wang2025tcim}. Table~\ref{tab:datasets-v11} lists the
full name, category, number of vertices $n$, and number of edges $m$ for each
network. For Flixster, we use the released edge probabilities learned
from user interaction logs and the accompanying ten sources.
For the SNAP networks, we use the same weighted cascade setting
as in prior work~\cite{xie2023vertex,sandimin2024}, assigning
$p_{uv}=1/d^-(v)$ in the original graph before blocking.
The graph files for \texttt{EA}, \texttt{YT}, \texttt{ST}, and
\texttt{TW} come from the data released with \texttt{AG} and
\texttt{GR}~\cite{xie2025blocking,xie2024artifact}.
For \texttt{YT}, we use the \texttt{AG}/\texttt{GR} repository version as
a directed graph.
For Twitter, we keep the first occurrence of each directed edge.
For each SNAP network, we use a fixed random seed to draw ten
sources without replacement from the 200 vertices with the highest
outdegree.
All methods use the same source set $S$ and candidate set $C$.

\begin{table}[t]
\caption{Datasets used in the experiments.}
\label{tab:datasets-v11}
\centering
\begingroup
\fontsize{8}{9.2}\selectfont
\setlength{\tabcolsep}{3pt}
\renewcommand{\arraystretch}{1.12}
\begin{tabular}{|c|c|c|c|c|}
\hline
Dataset & Full Name & $n$ & $m$ & Category \\
\hline
\texttt{GN} & \texttt{p2p-Gnutella08} & 6,301 & 20,777 & Peer-to-peer \\
\texttt{WV} & \texttt{wiki-Vote} & 7,115 & 103,689 & Social \\
\texttt{FS} & \texttt{Flixster} & 28,843 & 119,018 & Social \\
\texttt{EP} & \texttt{soc-Epinions1} & 75,879 & 508,837 & Social \\
\texttt{TW} & \texttt{ego-Twitter} & 81,306 & 1,768,149 & Social \\
\texttt{EA} & \texttt{email-EuAll} & 265,214 & 420,045 & Communication \\
\texttt{ST} & \texttt{web-Stanford} & 281,903 & 2,312,497 & Web \\
\texttt{YT} & \texttt{com-Youtube} & 1,134,890 & 2,987,624 & Social \\
\hline
\end{tabular}
\endgroup
\end{table}

\paragraph{Baselines}
We compare with the following original author implementations.
\begin{itemize}
\item \sand~\cite{sandimin2024}: Forms candidates from sampled lower
and upper bounds and a degree heuristic, then selects among them by
estimated residual influence.
\item \sandminus~\cite{sandimin2024}: The authors' variant that omits
the upper-bound candidate and compares the lower-bound candidate with
the degree heuristic.
\item \texttt{CBFM}~\cite{wang2025tcim}: Uses sampled common-path
lower bounds and live-path upper bounds. We evaluate its classical
IMIN branch under the same fixed-source objective.
\item \texttt{AG}~\cite{xie2025blocking}: AdvancedGreedy uses dominator
structures on sampled live-edge graphs to evaluate greedy blocking gains.
\item \texttt{GR}~\cite{xie2025blocking}: GreedyReplace supplements
greedy selection with blocker-replacement steps.
\end{itemize}
On Flixster, \texttt{AG} and \texttt{GR} use Bernoulli
draws with the supplied edge probabilities.

\paragraph{Setting}
We run all methods on a single thread on an Intel
Core i9-10850K with 31.87 GiB RAM, Windows 11, and GCC 12.1.0.
We use a runtime limit of 1,800~s and a memory limit of 8~GiB.
TL indicates a running time over 1,800~s.
We run each configuration three times with three random seeds and
measure total running time from start to finish. We report the mean
and bold the best value in each row.
Unless stated otherwise, we use $k\in\{20,100\}$. \method uses
$\rho=0.6$, $P_0=65{,}536$, and $J=5$.
For \sand and \sandminus, $\epsilon=0.2$ and $\gamma=\beta=0.1$.
For \texttt{CBFM}, $\epsilon=\gamma=\beta=0.1$.
\texttt{AG}/\texttt{GR} use $10^4$ live-edge samples per selection step;
\texttt{GR} uses $10^5$ trials per internal influence estimate.

\subsection{Efficiency on Large Networks}
\label{sec:eval-main}

We compare the running time of \method with that of the baselines at
blocker budgets $k=20$ and $k=100$.

\begin{table}[t]
\caption{Total running time (s).}
\label{tab:main-runtime-v11}
\centering
\begingroup
\fontsize{8}{9.2}\selectfont
\setlength{\tabcolsep}{2.6pt}
\renewcommand{\arraystretch}{1.08}
\begin{tabular}{@{}cccccccc@{}}
\toprule
Data & $k$ & \texttt{SIG} & \sandminus & \sand & \texttt{CBFM} & \texttt{AG} & \texttt{GR} \\
\midrule
\texttt{GN} & 20 & \textbf{1.08} & 2.60 & 4.29 & 1.59 & 21.50 & 19.80 \\
\texttt{GN} & 100 & \textbf{0.03} & 0.11 & 0.10 & 0.28 & 91.04 & 62.57 \\
\texttt{WV} & 20 & \textbf{0.72} & 2.05 & 7.04 & 1.95 & 30.60 & 29.49 \\
\texttt{WV} & 100 & \textbf{0.72} & 2.14 & 5.73 & 1.86 & 139.92 & 106.54 \\
\texttt{FS} & 20 & \textbf{2.45} & 10.98 & 22.98 & 11.18 & 160.59 & 149.58 \\
\texttt{FS} & 100 & \textbf{0.40} & 8.95 & 17.97 & 7.56 & 763.41 & 536.25 \\
\texttt{EP} & 20 & \textbf{3.90} & 49.24 & 200.49 & 44.34 & 559.33 & 545.93 \\
\texttt{EP} & 100 & \textbf{1.38} & 49.83 & 227.20 & 38.13 & TL & TL \\
\texttt{TW} & 20 & \textbf{11.66} & 58.83 & 360.77 & 39.12 & 643.38 & 693.34 \\
\texttt{TW} & 100 & \textbf{4.96} & 62.16 & 318.49 & 48.74 & TL & TL \\
\texttt{EA} & 20 & \textbf{10.32} & 809.66 & TL & 560.48 & TL & 1,718.34 \\
\texttt{EA} & 100 & \textbf{4.29} & 810.70 & TL & 571.17 & TL & TL \\
\texttt{ST} & 20 & \textbf{9.87} & TL & TL & 877.21 & TL & 1,753.80 \\
\texttt{ST} & 100 & \textbf{19.16} & TL & TL & 938.12 & TL & TL \\
\texttt{YT} & 20 & \textbf{2.31} & TL & TL & 1,254.26 & TL & TL \\
\texttt{YT} & 100 & \textbf{2.30} & TL & TL & 1,343.18 & TL & TL \\
\bottomrule
\end{tabular}
\endgroup
\end{table}

\method has the lowest mean total running time in every
comparison in Table~\ref{tab:main-runtime-v11}. It completes
all three repetitions on every graph at both budgets, as does
\texttt{CBFM}. The speedup over the fastest completing baseline ranges
from $1.47\times$ on \texttt{GN} at $k=20$ to $583\times$ on \texttt{YT}
at $k=100$. Across both budgets on \texttt{EA}, \texttt{ST}, and
\texttt{YT}, \texttt{CBFM} is the fastest completing baseline and
\method is about $49$--$583\times$ faster. On \texttt{YT} at $k=100$,
\method takes 2.30~s versus 1,343.18~s for \texttt{CBFM}. At the same
budget, \texttt{AG} and \texttt{GR} reach the 1,800~s limit, which is
over $780$ times \method's mean running time.

On \texttt{FS}, \method is $4.49\times$ and
$18.86\times$ faster than the fastest completing baseline at
$k=20$ and $k=100$, respectively. On \texttt{TW}, the corresponding
speedups are $3.35\times$ and $9.82\times$. At $k=100$ on
\texttt{TW}, \method takes 4.96~s and the fastest baseline,
\texttt{CBFM}, takes 48.74~s. \texttt{AG} and \texttt{GR} reach the
1,800~s limit at that budget.

The number of sampled stages helps explain how the blocker budget
affects running time. On \texttt{EP}, increasing $k$ from 20 to 100 reduces
the number of sampled stages from two to one in every repetition, and
the mean running time falls from 3.90~s to 1.38~s. On \texttt{ST}, the
stage count instead rises from four to five, and the time increases
from 9.87~s to 19.16~s. There is also an exact structural case on
\texttt{GN} at $k=100$. The 96 vertices that can be directly activated
from the sources fit in the budget, so \method returns that set without
sampling. Its mean running time in this case is 0.03~s.
These results show how \method saves time by certifying in fewer
sampled stages or, when the structure permits, returning an exact
solution directly.

Across the tested settings, \method is faster than every baseline, with
speedups of up to $583\times$ over the fastest completing baseline.
It finishes in under 20~s at both budgets on the four largest graphs,
including one with over 1.1 million vertices and 2.9 million edges.

\subsection{Quality Validation}
\label{sec:eval-quality}

\begin{table}[t]
\caption{Influence reduction $D$.}
\label{tab:quality-v11}
\centering
\begingroup
\fontsize{8}{9.2}\selectfont
\setlength{\tabcolsep}{2.5pt}
\renewcommand{\arraystretch}{1.08}
\begin{tabular}{@{}cccccccc@{}}
\toprule
Data & $k$ & \texttt{SIG} & \sandminus & \sand & \texttt{CBFM} & \texttt{AG} & \texttt{GR} \\
\midrule
\texttt{GN} & 20 & 205.90 & 204.36 & 204.28 & 203.65 & 205.90 & \textbf{206.03} \\
\texttt{GN} & 100 & \textbf{233.44} & \textbf{233.44} & \textbf{233.44} & \textbf{233.44} & \textbf{233.44} & \textbf{233.44} \\
\texttt{WV} & 20 & 20.50 & 20.39 & 19.87 & 19.80 & 20.51 & \textbf{20.52} \\
\texttt{WV} & 100 & 47.14 & 46.74 & 46.90 & 44.17 & 47.18 & \textbf{47.18} \\
\texttt{FS} & 20 & 270.60 & 158.48 & 196.49 & 269.20 & 274.48 & \textbf{275.33} \\
\texttt{FS} & 100 & 537.43 & 394.57 & 394.57 & 523.03 & \textbf{553.10} & 552.28 \\
\texttt{EP} & 20 & 354.25 & 341.60 & 334.71 & 348.07 & \textbf{354.36} & 329.06 \\
\texttt{EP} & 100 & 724.08 & 734.16 & \textbf{734.92} & 697.36 & TL & TL \\
\texttt{TW} & 20 & 421.04 & 422.04 & 413.88 & 402.29 & \textbf{422.90} & 422.76 \\
\texttt{TW} & 100 & 882.75 & 891.38 & \textbf{892.62} & 868.94 & TL & TL \\
\texttt{EA} & 20 & 175.10 & \textbf{175.58} & TL & 170.43 & TL & 173.96 \\
\texttt{EA} & 100 & 415.63 & \textbf{420.33} & TL & 392.27 & TL & TL \\
\texttt{ST} & 20 & \textbf{322.98} & TL & TL & 318.21 & TL & 318.14 \\
\texttt{ST} & 100 & \textbf{421.93} & TL & TL & 403.81 & TL & TL \\
\texttt{YT} & 20 & \textbf{18,887.86} & TL & TL & 18,601.38 & TL & TL \\
\texttt{YT} & 100 & \textbf{35,805.76} & TL & TL & 34,878.63 & TL & TL \\
\bottomrule
\end{tabular}
\endgroup
\end{table}

\textcolor{black}{We first verify and confirm that in every experimental run,
\method returns a solution certified to meet the requested approximation ratio.
We then compare how much influence \method and the
baselines prevent with the blocker sets they return. \textcolor{black}{We evaluate all solutions
using 1,048,576 shared reverse-target samples per graph, drawn independently
of selection.} Table~\ref{tab:quality-v11}
reports the estimated number of activations prevented, averaged over
three runs.}

\textcolor{black}{Against \texttt{AG} and \texttt{GR}, \method stays
close to the better greedy result wherever either baseline completes.
On \texttt{GN}, \texttt{WV}, \texttt{EP}, and \texttt{TW}, the gap
is within 0.5\% at each such budget. The largest gap occurs on
\texttt{FS} at $k=100$, where \method prevents an estimated 537.43
activations, compared with 553.10 for \texttt{AG} and 552.28 for
\texttt{GR}. At $k=100$, both greedy baselines reach the time limit
on the four largest networks, while \method completes on all four.
Against \sandminus{} and \sand, \method makes its largest gains on
\texttt{FS}. At $k=20$, it prevents an estimated 270.60 activations,
70.7\% more than \sandminus{} and 37.7\% more than \sand.
At $k=100$, it prevents 36.2\% more than either variant.
On \texttt{EP} at $k=20$, \method leads the two variants by 3.7\%
and 5.8\%, respectively. At $k=100$ on the same graph, \method
trails both variants by less than 1.5\%.
\method outperforms \texttt{CBFM} in 15 settings and ties on
\texttt{GN} at $k=100$. On the two largest graphs, \texttt{ST} and
\texttt{YT}, \texttt{CBFM} is the only baseline to complete at
$k=100$. \method prevents 4.49\% more activations on \texttt{ST}
(421.93 vs.\ 403.81) and 2.66\% more on \texttt{YT}
(35,805.76 vs.\ 34,878.63).}

\textcolor{black}{Overall, \method reaches at least 97.1\% of the highest estimated
influence reduction in every main comparison, matching or exceeding
the best baseline result in multiple settings. It also returns
solutions in tests where several baselines exceed the runtime limit.}

\setlength{\textfloatsep}{10pt plus 2pt minus 2pt}
\setlength{\floatsep}{6pt plus 2pt minus 2pt}
\subsection{Scalability with Data Size}
\label{sec:eval-scalability}

\textcolor{black}{We test how \method and the baselines scale as the
graph grows at a fixed blocker budget. On \texttt{EA} and \texttt{ST},
we construct nested induced subgraphs by retaining 10, 25, 50, 75, or
100\% of the vertices outside $S$ in a fixed order. Each subgraph
includes all sources and keeps the original propagation probabilities
of its edges. We fix $k=20$ and compare mean total running times in
Figure~\ref{fig:scalability-v10}. Dashed arrows extend beyond the
1,800~s line at the first observed timeout on each curve.}

\textcolor{black}{\method is consistently the fastest across both
graphs and all five sizes, with speedups ranging from about
$17\times$ to $134\times$ over the fastest completing baseline.
At full size, \method takes 10.26~s on \texttt{EA} and 9.36~s on
\texttt{ST}, about $65\times$ and $96\times$ faster than \texttt{CBFM},
respectively. In contrast, \sand reaches the time limit on full
\texttt{EA} and at 50\% on \texttt{ST}, and \sandminus reaches it on
full \texttt{ST}.}

\textcolor{black}{The growth in \method's running time remains modest
at larger sizes. From 50\% to 100\%, the edge count nearly quadruples
on both graphs, while \method stays between 6.92 and 10.26~s on
\texttt{EA} and between 6.01 and 9.36~s on \texttt{ST}. \method
therefore retains its runtime advantage as the graph grows to more
than 2.3 million edges.}

\begin{figure}[t]
\setlength{\abovecaptionskip}{2pt}
\centering
\includegraphics[width=0.70\columnwidth,trim={0bp 13.5bp 0bp 6.0bp},clip]{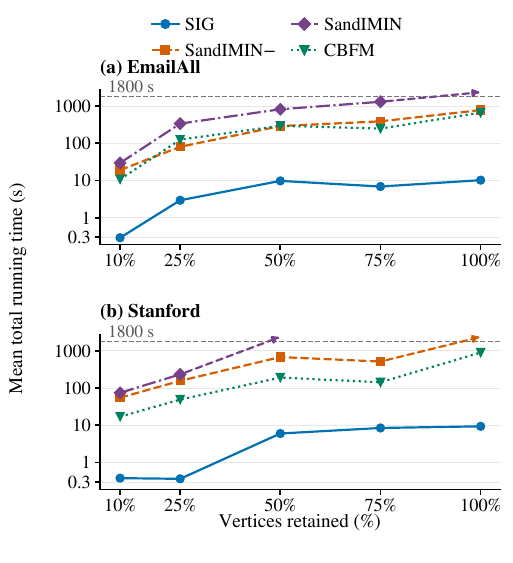}
\caption{\textcolor{black}{Mean total running time for vertex scalability.}}
\label{fig:scalability-v10}
\end{figure}

\subsection{Scalability with Budget Limit}
\label{sec:eval-budget-scalability}

\textcolor{black}{We study how allowing larger blocker budgets affects
the runtime and effectiveness of \method. We use the four largest
networks, \texttt{TW}, \texttt{EA}, \texttt{ST}, and \texttt{YT},
and vary $k$ over $\{10,20,50,100,200,500,1000\}$.
The graph, sources, and propagation probabilities stay fixed.
We run each budget three times and compare the mean running time
and influence reduction in Figures~\ref{fig:budget-runtime-v1} and~\ref{fig:budget-quality-v2}.}

\textcolor{black}{Figure~\ref{fig:budget-runtime-v1} shows that \method
maintains low running times as the budget grows by two orders of
magnitude. All means are at most 21.83~s, and the \texttt{YT} curve
stays between 2.24~s and 2.35~s. From $k=10$ to $k=1000$, mean time
falls from 12.69~s to 2.22~s on \texttt{TW}, from 21.83~s to 10.44~s
on \texttt{EA}, and from 19.65~s to 5.27~s on \texttt{ST}.
These decreases coincide with fewer adaptive stages: all three
\texttt{TW} runs need three stages at $k=10$ and one at $k=1000$,
while the corresponding \texttt{ST} runs use five and three.
Thus, selecting more blockers need not require more sampling.
\method can certify a larger set with fewer stages, offsetting the
extra greedy choices and keeping the running time low.}

\begin{figure}[t]
\setlength{\abovecaptionskip}{2pt}
\centering
\includegraphics[width=0.70\columnwidth,trim={0bp 4.5bp 0bp 8.5bp},clip]{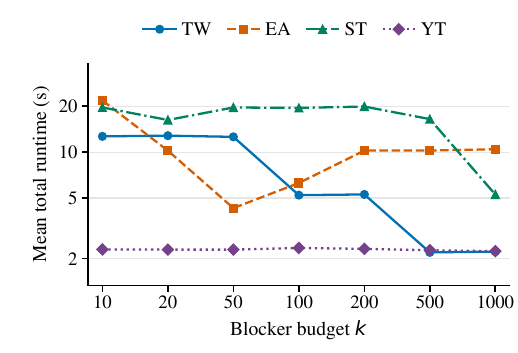}
\caption{\textcolor{black}{Mean total running time of SIG as the blocker budget varies.}}
\label{fig:budget-runtime-v1}
\end{figure}

\textcolor{black}{Figure~\ref{fig:budget-quality-v2} shows that \method
makes effective use of larger blocking budgets: mean influence
reduction increases at every tested budget on all four networks.
Raising $k$ from 10 to 1000 increases the mean reduction by a factor
of $5.36$ on \texttt{TW}, $11.97$ on \texttt{EA}, $3.18$ on
\texttt{ST}, and $4.16$ on \texttt{YT}. The gains continue beyond
$k=100$, with a further $1.56$--$3.25$-fold increase by $k=1000$.
Together, the two figures show that \method can deliver substantially
stronger blocking solutions as the budget grows while maintaining
consistently low running times on large networks.}

\begin{figure}[t]
\setlength{\abovecaptionskip}{2pt}
\centering
\includegraphics[width=0.70\columnwidth,trim={0bp 4.5bp 0bp 7.0bp},clip]{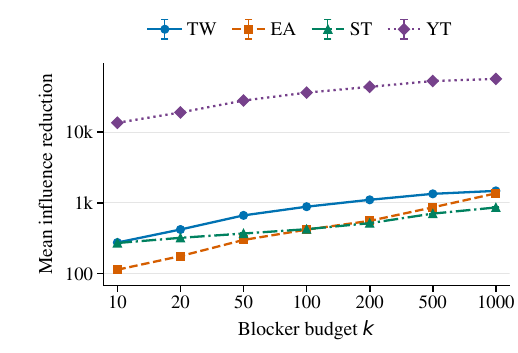}
\caption{\textcolor{black}{Mean influence reduction of SIG as the blocker budget varies.}}
\label{fig:budget-quality-v2}
\end{figure}

\subsection{Effect of the Parameter}
\label{sec:eval-parameters}

\textcolor{black}{We test whether \method's blocker quality is sensitive
to its requested approximation ratio $\rho$ or initial sample size
$P_0$. At $k=20$ on \texttt{EA} and \texttt{YT}, we vary one parameter
at a time around the defaults $\rho=0.6$ and $P_0=2^{16}$.
Across the tested choices, estimated influence reduction stays within
2.52\% of the default on \texttt{EA} and within 0.13\% on \texttt{YT}.}

\textcolor{black}{A stricter guarantee mainly increases the work needed
for certification. Raising $\rho$ from 0.6 to 0.8 changes estimated
influence reduction by $-0.02$ on \texttt{EA} and $+16.11$ on
\texttt{YT}, with paired Monte Carlo standard errors of 0.29 and
2.84, respectively. At $\rho=0.8$, every run uses all five stages,
and mean running time rises from 10.20 to 45.70~s on \texttt{EA}
and from 2.22 to 41.77~s on \texttt{YT}
(Figure~\ref{fig:parameters-rho}).}

\textcolor{black}{The initial sample size determines how much work
\method spends up front. On \texttt{EA}, increasing $P_0$ from
$2^{14}$ to $2^{18}$ reduces the number of stages from five to one
and mean running time from 11.44 to 6.04~s. On \texttt{YT}, the
default already certifies in one stage, so increasing $P_0$ to
$2^{18}$ raises mean running time from 2.22 to 6.05~s
(Figure~\ref{fig:parameters-p0}). These results show that \method
maintains similar blocker quality across a range of parameter choices,
while the default achieves that quality without the cost of the
strictest certification target.}

\begin{figure}[t]
\setlength{\abovecaptionskip}{2pt}
\centering
\includegraphics[width=0.70\columnwidth,trim={0bp 11.0bp 0bp 7.0bp},clip]{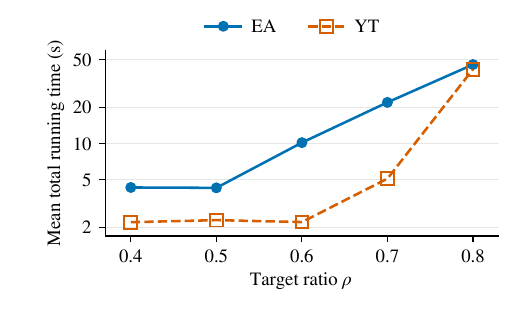}
\caption{\textcolor{black}{Mean total running time as the requested approximation ratio $\rho$ varies.}}
\label{fig:parameters-rho}
\par\vspace{1pt}
\includegraphics[width=0.70\columnwidth,trim={0bp 11.0bp 0bp 7.0bp},clip]{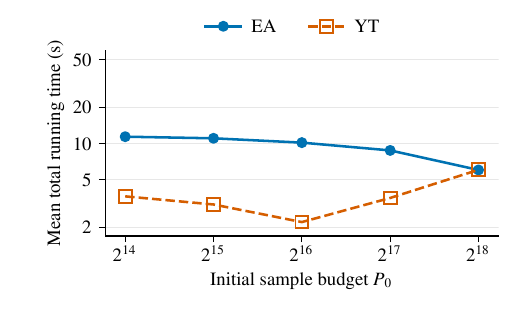}
\caption{\textcolor{black}{Mean total running time as the initial sample size $P_0$ varies.}}
\label{fig:parameters-p0}
\end{figure}

\subsection{Ablation Study}
\label{sec:eval-ablation}

\textcolor{black}{We test how source integration and the validation
bounds affect running time and solution quality. We compare the full
\method with three variants on \texttt{EP}, \texttt{EA}, \texttt{ST},
and \texttt{YT} at $k=100$, using matched inputs and seeds.}

\begin{itemize}
\item \textcolor{black}{\texttt{SIG-Bernoulli}. Samples a source outcome
in each training realization in place of conditional source integration.
We compare the first-stage blocker sets at equal training sample counts
and the final sets at termination.}
\item \textcolor{black}{\texttt{SIG-Chernoff}. Removes second-moment
computation and empirical Bernstein bounds from validation. We compare
the work needed to certify the same approximation target.}
\item \textcolor{black}{\texttt{SIG-Endpoint}. Uses only weights 0 and 1
in both moment upper bounds. We test whether the interior weights in
\method allow it to certify sooner.}
\end{itemize}

\textcolor{black}{Figure~\ref{fig:ablation-efficiency-k100} shows that
\method is $2.67\times$ to $2.77\times$ faster than the three variants
on \texttt{EP} and $1.90\times$ to $2.30\times$ faster on \texttt{EA}.
Compared with \texttt{SIG-Chernoff}, \method reduces the number of
validation samples from 196,608 to 65,536 on \texttt{EP} and from
458,752 to 196,608 on \texttt{EA}. On \texttt{ST} and \texttt{YT},
all four methods stop at the same stage, and their running times are
within 0.35~s of one another on each graph.}

\begin{figure}[t]
\setlength{\abovecaptionskip}{2pt}
\centering
\includegraphics[width=0.76\columnwidth,trim={0bp 8.5bp 0bp 5.5bp},clip]{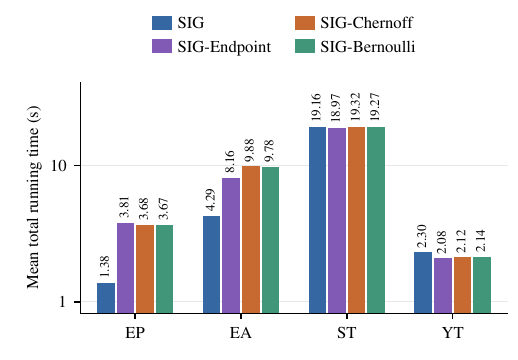}
\caption{\textcolor{black}{Mean total running time in the ablation study at $k=100$.}}
\label{fig:ablation-efficiency-k100}
\end{figure}

\textcolor{black}{\texttt{SIG-Endpoint} and \texttt{SIG-Chernoff}
achieve similar final influence reduction to \method. On \texttt{EP}
and \texttt{EA}, their estimates differ from the paired \method
estimates by at most 1.32\%. On \texttt{ST} and \texttt{YT}, both
variants return the same blocker sets as \method in every repetition.
\texttt{SIG-Bernoulli} yields lower influence reduction on all four
graphs. At equal first-stage training sample counts, the decrease is
2.93\% to 10.00\%, and at termination it is 1.40\% to 4.65\%.}

\textcolor{black}{The full \method offers the strongest overall balance
of speed and solution quality across the four graphs. Source integration
improves blocker quality consistently, and the full validation bounds
cut certification work on \texttt{EP} and \texttt{EA} while keeping
influence reduction close to that of the two validation variants.}

\section{Conclusion}
\label{sec:conclusion}

{\color{black}We \textcolor{black}{present} \method, a sampling framework for influence
minimization by node blocking under the IC model.
\textcolor{black}{It groups source outcomes by the vertices that can individually
protect a target and maintains their probabilities in compact weighted
trees. We prove that its sampling mechanism gives unbiased estimates
of the lower-bound objective without increasing variance for fixed
blocking plans, and establish high-probability bounds that validate
its reported approximation certificates.}
\textcolor{black}{Extensive experiments show that \method has the lowest mean total
running time in all main comparisons, with speedups of up to
$583\times$ over the fastest baseline. \method produces solutions competitive with those of prior methods and
achieves the highest influence reduction in multiple settings,
including on many of the largest networks, where several baselines exceed
the runtime limit.}
\textcolor{black}{Additional experiments show that \method maintains consistently
low running times as graph size and budget increase, and delivers
stronger blocking solutions with larger budgets. Parameter and
ablation studies assess sensitivity and the effectiveness of its
sampling and validation designs.}
\textcolor{black}{A possible future direction is to extend \method to evolving networks
by updating its weighted samples as edges and propagation
probabilities change.}}

\clearpage
\endgroup
\section*{Acknowledgment of AI-Generated Content}
\textcolor{black}{The authors acknowledge the use of OpenAI Codex to assist with literature
review, implementation of SIG, and polishing of the abstract and Sections
I, II, and IV--VII. Under the authors' direction,
Codex ran and managed the experiments reported in Section VI, processed and
checked the results, prepared the figures and tables for that section, and
organized the public artifact repository containing the source code and
experiment scripts. The authors led the mathematical derivations and proofs
in Sections IV and V, with Codex suggesting approaches and helping check
and clarify the arguments. The authors take full responsibility for the
mathematical claims and proofs, the experimental results and their
interpretation, and all other content of this paper.}
\bibliographystyle{IEEEtran}
\begingroup
\makeatletter
\def\review@newkeys{tang2014tim,tang2015imm,nguyen2016ssa,huang2017revisiting,chen2015topic,ohsaka2016dynamic,arora2017debunking,han2018adaptive,guo2020rr,bian2020budgeted,tsaras2021collective,huang2022closed,huang2023stakeholders,guo2023budgetadaptive,feng2024seedselection,chen2009efficient,chen2010scalable,chen2016robust,he2016robust,sun2018multiround,wu2019factorization,zhang2020geodemographic,zhang2023capacity,feng2023fairness,chen2025link,cheng2014imrank,zhang2017game,xiang2025hypermin,chang2025timin,shi2019adaptive,tang2022distributed,zhao2019tracking,huang2019temporal,li2019multifaceted,khan2016revenue,huang2020profit,wang2023parallel,xie2024countering,huang2024matroid,borgs2014rr,goyal2010probabilities,jin2011distance,ke2019reliability,teng2023iminimize,mccabe2024deplatforming}
\@for\review@key:=\review@newkeys\do{%
  \expandafter\def\csname review@bluebib@\review@key\endcsname{}%
}
\def\review@purplekey{chen2020lattice}
\let\review@originalbibitem\@bibitem
\def\@bibitem#1{%
  \ifcsname review@bluebib@#1\endcsname
    \color{black}%
  \else
    \color{black}%
  \fi
  \def\review@currentkey{#1}%
  \ifx\review@currentkey\review@purplekey
    \color{black}%
  \fi
  \review@originalbibitem{#1}%
}
\makeatother
\bibliography{references}

@inproceedings{kempe2003influence,
  author = {Kempe, David and Kleinberg, Jon and Tardos, {\'E}va},
  title = {Maximizing the Spread of Influence through a Social Network},
  booktitle = {Proceedings of the Ninth ACM SIGKDD International Conference on Knowledge Discovery and Data Mining},
  pages = {137--146},
  year = {2003},
  doi = {10.1145/956750.956769},
  url = {https://doi.org/10.1145/956750.956769}
}

@article{blackwell1947,
  author = {Blackwell, David},
  title = {Conditional Expectation and Unbiased Sequential Estimation},
  journal = {The Annals of Mathematical Statistics},
  volume = {18},
  number = {1},
  pages = {105--110},
  year = {1947},
  doi = {10.1214/aoms/1177730497},
  url = {https://doi.org/10.1214/aoms/1177730497}
}

@IEEEtranBSTCTL{icdebstcontrol,
  CTLdash_repeated_names = {no}
}

@misc{snapnets,
  author = {Jure Leskovec and Andrej Krevl},
  title = {{SNAP Datasets}: {Stanford} Large Network Dataset Collection},
  howpublished = {\url{http://snap.stanford.edu/data}},
  month = jun,
  year = {2014}
}

@article{sandimin2026,
  author = {Wang, Jinghao and Wu, Yanping and Wang, Xiaoyang and Zhang, Ying and Qin, Lu and Zhang, Wenjie and Lin, Xuemin},
  title = {Efficient and cost-effective influence and blocker minimization},
  journal = {The VLDB Journal},
  year = {2026},
  volume = {35},
  number = {6},
  month = nov,
  note = {Article 53},
  doi = {10.1007/s00778-026-01003-4},
  url = {https://doi.org/10.1007/s00778-026-01003-4}
}

@article{xie2025blocking,
  author = {Xie, Jiadong and Zhang, Fan and Wang, Kai and Liu, Jialu and Lin, Xuemin and Zhang, Wenjie},
  title = {Influence Minimization via Blocking Strategies},
  journal = {INFORMS Journal on Computing},
  volume = {37},
  number = {6},
  pages = {1587--1604},
  year = {2025},
  month = nov,
  doi = {10.1287/ijoc.2024.0591},
  url = {https://doi.org/10.1287/ijoc.2024.0591}
}

@misc{xie2024artifact,
  author = {Xie, Jiadong and Zhang, Fan and Wang, Kai and Liu, Jialu and Lin, Xuemin and Zhang, Wenjie},
  title = {Influence Minimization via Blocking Strategies},
  year = {2024},
  note = {Software and data, version 16daafb0af360e932454791caade5b97779a2347},
  url = {https://github.com/INFORMSJoC/2024.0591}
}

@inproceedings{xie2023vertex,
  author = {Xie, Jiadong and Zhang, Fan and Wang, Kai and Lin, Xuemin and Zhang, Wenjie},
  title = {Minimizing the Influence of Misinformation via Vertex Blocking},
  booktitle = {2023 IEEE 39th International Conference on Data Engineering (ICDE)},
  pages = {789--801},
  year = {2023},
  doi = {10.1109/ICDE55515.2023.00066},
  url = {https://doi.org/10.1109/ICDE55515.2023.00066}
}

@inproceedings{maurer2009,
  author = {Maurer, Andreas and Pontil, Massimiliano},
  title = {Empirical Bernstein Bounds and Sample Variance Penalization},
  booktitle = {Proceedings of the Conference on Learning Theory},
  year = {2009},
  url = {https://www.cs.mcgill.ca/~colt2009/papers/012.pdf}
}

@article{sandimin2024,
  author  = {Wang, Jinghao and Wu, Yanping and Wang, Xiaoyang and Zhang, Ying and Qin, Lu and Zhang, Wenjie and Lin, Xuemin},
  title   = {Efficient Influence Minimization via Node Blocking},
  journal = {Proceedings of the VLDB Endowment},
  year    = {2024},
  volume  = {17},
  number  = {10},
  pages   = {2501--2513},
  doi     = {10.14778/3675034.3675042},
  url     = {https://www.vldb.org/pvldb/vol17/p2501-wu.pdf}
}

@article{chen2020lattice,
  author  = {Chen, Wei and Wu, Ruihan and Yu, Zheng},
  title   = {Scalable Lattice Influence Maximization},
  journal = {IEEE Transactions on Computational Social Systems},
  year    = {2020},
  month   = aug,
  volume  = {7},
  number  = {4},
  pages   = {956--970},
  doi     = {10.1109/TCSS.2020.2998777},
  url     = {https://doi.org/10.1109/TCSS.2020.2998777}
}

@article{pham2019blocking,
  author  = {Pham, Canh V. and Phu, Quat V. and Hoang, Huan X. and Pei, Jun and Thai, My T.},
  title   = {Minimum budget for misinformation blocking in online social networks},
  journal = {Journal of Combinatorial Optimization},
  year    = {2019},
  month   = nov,
  volume  = {38},
  number  = {4},
  pages   = {1101--1127},
  doi     = {10.1007/s10878-019-00439-5},
  url     = {https://doi.org/10.1007/s10878-019-00439-5}
}

@article{nemhauser1978,
  author  = {Nemhauser, G. L. and Wolsey, L. A. and Fisher, M. L.},
  title   = {An analysis of approximations for maximizing submodular set functions---{I}},
  journal = {Mathematical Programming},
  year    = {1978},
  month   = dec,
  volume  = {14},
  number  = {1},
  pages   = {265--294},
  doi     = {10.1007/BF01588971},
  url     = {https://doi.org/10.1007/BF01588971}
}

@article{wang2025tcim,
  author = {Wang, Jinghao and Wu, Yanping and Wang, Xiaoyang and Zhang, Ying and Zhang, Wenjie and Qin, Lu},
  title = {Time-Critical Influence Minimization via Node Blocking},
  journal = {Proceedings of the ACM on Management of Data},
  volume = {3},
  number = {6},
  year = {2025},
  month = dec,
  pages = {369:1--369:27},
  doi = {10.1145/3769834},
  url = {https://doi.org/10.1145/3769834}
}

@inproceedings{iyer2013semidifferential,
  author = {Iyer, Rishabh and Jegelka, Stefanie and Bilmes, Jeff},
  title = {Fast Semidifferential-based Submodular Function Optimization},
  booktitle = {Proceedings of the 30th International Conference on Machine Learning},
  series = {Proceedings of Machine Learning Research},
  volume = {28},
  number = {3},
  pages = {855--863},
  year = {2013},
  url = {https://proceedings.mlr.press/v28/iyer13.html}
}

@inproceedings{tang2014tim,
  author = {Tang, Youze and Xiao, Xiaokui and Shi, Yanchen},
  title = {Influence Maximization: Near-Optimal Time Complexity Meets Practical Efficiency},
  booktitle = {Proceedings of the 2014 ACM SIGMOD International Conference on Management of Data},
  pages = {75--86},
  year = {2014},
  doi = {10.1145/2588555.2593670},
  url = {https://doi.org/10.1145/2588555.2593670}
}

@inproceedings{tang2015imm,
  author = {Tang, Youze and Shi, Yanchen and Xiao, Xiaokui},
  title = {Influence Maximization in Near-Linear Time: A Martingale Approach},
  booktitle = {Proceedings of the 2015 ACM SIGMOD International Conference on Management of Data},
  pages = {1539--1554},
  year = {2015},
  doi = {10.1145/2723372.2723734},
  url = {https://doi.org/10.1145/2723372.2723734}
}

@inproceedings{nguyen2016ssa,
  author = {Nguyen, Hung T. and Thai, My T. and Dinh, Thang N.},
  title = {Stop-and-Stare: Optimal Sampling Algorithms for Viral Marketing in Billion-scale Networks},
  booktitle = {Proceedings of the 2016 International Conference on Management of Data},
  pages = {695--710},
  year = {2016},
  doi = {10.1145/2882903.2915207},
  url = {https://doi.org/10.1145/2882903.2915207}
}

@article{huang2017revisiting,
  author = {Huang, Keke and Wang, Sibo and Bevilacqua, Glenn and Xiao, Xiaokui and Lakshmanan, Laks V. S.},
  title = {Revisiting the Stop-and-Stare Algorithms for Influence Maximization},
  journal = {Proceedings of the VLDB Endowment},
  volume = {10},
  number = {9},
  pages = {913--924},
  year = {2017},
  doi = {10.14778/3099622.3099623},
  url = {https://doi.org/10.14778/3099622.3099623}
}

@article{chen2015topic,
  author = {Chen, Shuo and Fan, Ju and Li, Guoliang and Feng, Jianhua and Tan, Kian-lee and Tang, Jinhui},
  title = {Online Topic-Aware Influence Maximization},
  journal = {Proceedings of the VLDB Endowment},
  volume = {8},
  number = {6},
  pages = {666--677},
  year = {2015},
  doi = {10.14778/2735703.2735706},
  url = {https://doi.org/10.14778/2735703.2735706}
}

@article{ohsaka2016dynamic,
  author = {Ohsaka, Naoto and Akiba, Takuya and Yoshida, Yuichi and Kawarabayashi, Ken-ichi},
  title = {Dynamic Influence Analysis in Evolving Networks},
  journal = {Proceedings of the VLDB Endowment},
  volume = {9},
  number = {12},
  pages = {1077--1088},
  year = {2016},
  doi = {10.14778/2994509.2994525},
  url = {https://doi.org/10.14778/2994509.2994525}
}

@article{han2018adaptive,
  author = {Han, Kai and Huang, Keke and Xiao, Xiaokui and Tang, Jing and Sun, Aixin and Tang, Xueyan},
  title = {Efficient Algorithms for Adaptive Influence Maximization},
  journal = {Proceedings of the VLDB Endowment},
  volume = {11},
  number = {9},
  pages = {1029--1040},
  year = {2018},
  doi = {10.14778/3213880.3213883},
  url = {https://doi.org/10.14778/3213880.3213883}
}

@inproceedings{guo2020rr,
  author = {Guo, Qintian and Wang, Sibo and Wei, Zhewei and Chen, Ming},
  title = {Influence Maximization Revisited: Efficient Reverse Reachable Set Generation with Bound Tightened},
  booktitle = {Proceedings of the 2020 ACM SIGMOD International Conference on Management of Data},
  pages = {2167--2181},
  year = {2020},
  doi = {10.1145/3318464.3389740},
  url = {https://doi.org/10.1145/3318464.3389740}
}

@article{bian2020budgeted,
  author = {Bian, Song and Guo, Qintian and Wang, Sibo and Yu, Jeffrey Xu},
  title = {Efficient Algorithms for Budgeted Influence Maximization on Massive Social Networks},
  journal = {Proceedings of the VLDB Endowment},
  volume = {13},
  number = {9},
  pages = {1498--1510},
  year = {2020},
  doi = {10.14778/3397230.3397244},
  url = {https://doi.org/10.14778/3397230.3397244}
}

@article{tsaras2021collective,
  author = {Tsaras, Dimitris and Trimponias, George and Ntaflos, Lefteris and Papadias, Dimitris},
  title = {Collective Influence Maximization for Multiple Competing Products with an Awareness-to-Influence Model},
  journal = {Proceedings of the VLDB Endowment},
  volume = {14},
  number = {7},
  pages = {1124--1136},
  year = {2021},
  doi = {10.14778/3450980.3450981},
  url = {https://doi.org/10.14778/3450980.3450981}
}

@article{huang2022closed,
  author = {Huang, Shixun and Lin, Wenqing and Bao, Zhifeng and Sun, Jiachen},
  title = {Influence Maximization in Real-World Closed Social Networks},
  journal = {Proceedings of the VLDB Endowment},
  volume = {16},
  number = {2},
  pages = {180--192},
  year = {2022},
  doi = {10.14778/3565816.3565821},
  url = {https://doi.org/10.14778/3565816.3565821}
}

@article{huang2023stakeholders,
  author = {Huang, Shixun and Gan, Junhao and Bao, Zhifeng and Lin, Wenqing},
  title = {Managing Conflicting Interests of Stakeholders in Influencer Marketing},
  journal = {Proceedings of the ACM on Management of Data},
  volume = {1},
  number = {1},
  pages = {80:1--80:27},
  year = {2023},
  doi = {10.1145/3588934},
  url = {https://doi.org/10.1145/3588934}
}

@article{guo2023budgetadaptive,
  author = {Guo, Qintian and Feng, Chen and Zhang, Fangyuan and Wang, Sibo},
  title = {Efficient Algorithm for Budgeted Adaptive Influence Maximization: An Incremental {RR}-set Update Approach},
  journal = {Proceedings of the ACM on Management of Data},
  volume = {1},
  number = {3},
  pages = {207:1--207:26},
  year = {2023},
  doi = {10.1145/3617328},
  url = {https://doi.org/10.1145/3617328}
}

@article{feng2024seedselection,
  author = {Feng, Chen and Chen, Xingguang and Guo, Qintian and Zhang, Fangyuan and Wang, Sibo},
  title = {Efficient Approximation Algorithms for Minimum Cost Seed Selection with Probabilistic Coverage Guarantee},
  journal = {Proceedings of the ACM on Management of Data},
  volume = {2},
  number = {4},
  pages = {197:1--197:26},
  year = {2024},
  doi = {10.1145/3677133},
  url = {https://doi.org/10.1145/3677133}
}

@inproceedings{chen2016robust,
  author = {Chen, Wei and Lin, Tian and Tan, Zihan and Zhao, Mingfei and Zhou, Xuren},
  title = {Robust Influence Maximization},
  booktitle = {Proceedings of the 22nd ACM SIGKDD International Conference on Knowledge Discovery and Data Mining},
  pages = {795--804},
  year = {2016},
  doi = {10.1145/2939672.2939745},
  url = {https://doi.org/10.1145/2939672.2939745}
}

@inproceedings{he2016robust,
  author = {He, Xinran and Kempe, David},
  title = {Robust Influence Maximization},
  booktitle = {Proceedings of the 22nd ACM SIGKDD International Conference on Knowledge Discovery and Data Mining},
  pages = {885--894},
  year = {2016},
  doi = {10.1145/2939672.2939760},
  url = {https://doi.org/10.1145/2939672.2939760}
}

@inproceedings{sun2018multiround,
  author = {Sun, Lichao and Huang, Weiran and Yu, Philip S. and Chen, Wei},
  title = {Multi-Round Influence Maximization},
  booktitle = {Proceedings of the 24th ACM SIGKDD International Conference on Knowledge Discovery \& Data Mining},
  pages = {2249--2258},
  year = {2018},
  doi = {10.1145/3219819.3220101},
  url = {https://doi.org/10.1145/3219819.3220101}
}

@inproceedings{wu2019factorization,
  author = {Wu, Qingyun and Li, Zhige and Wang, Huazheng and Chen, Wei and Wang, Hongning},
  title = {Factorization Bandits for Online Influence Maximization},
  booktitle = {Proceedings of the 25th ACM SIGKDD International Conference on Knowledge Discovery \& Data Mining},
  pages = {636--646},
  year = {2019},
  doi = {10.1145/3292500.3330874},
  url = {https://doi.org/10.1145/3292500.3330874}
}

@inproceedings{zhang2020geodemographic,
  author = {Zhang, Kaichen and Zhou, Jingbo and Tao, Donglai and Karras, Panagiotis and Li, Qing and Xiong, Hui},
  title = {Geodemographic Influence Maximization},
  booktitle = {Proceedings of the 26th ACM SIGKDD International Conference on Knowledge Discovery \& Data Mining},
  pages = {2764--2774},
  year = {2020},
  doi = {10.1145/3394486.3403327},
  url = {https://doi.org/10.1145/3394486.3403327}
}

@inproceedings{zhang2023capacity,
  author = {Zhang, Shiqi and Huang, Yiqian and Sun, Jiachen and Lin, Wenqing and Xiao, Xiaokui and Tang, Bo},
  title = {Capacity Constrained Influence Maximization in Social Networks},
  booktitle = {Proceedings of the 29th ACM SIGKDD Conference on Knowledge Discovery and Data Mining},
  pages = {3376--3385},
  year = {2023},
  doi = {10.1145/3580305.3599267},
  url = {https://doi.org/10.1145/3580305.3599267}
}

@inproceedings{feng2023fairness,
  author = {Feng, Yuting and Patel, Ankitkumar and Cautis, Bogdan and Vahabi, Hossein},
  title = {Influence Maximization with Fairness at Scale},
  booktitle = {Proceedings of the 29th ACM SIGKDD Conference on Knowledge Discovery and Data Mining},
  pages = {4046--4055},
  year = {2023},
  doi = {10.1145/3580305.3599847},
  url = {https://doi.org/10.1145/3580305.3599847}
}

@inproceedings{chen2025link,
  author = {Chen, Xiaolong and Tang, Jing},
  title = {Scalable Link Recommendation for Influence Maximization},
  booktitle = {Proceedings of the 31st ACM SIGKDD Conference on Knowledge Discovery and Data Mining V.1},
  pages = {130--141},
  year = {2025},
  doi = {10.1145/3690624.3709190},
  url = {https://doi.org/10.1145/3690624.3709190}
}

@inproceedings{zhang2017game,
  author = {Zhang, Yu and Zhang, Yan},
  title = {Top-{K} Influential Nodes in Social Networks: A Game Perspective},
  booktitle = {Proceedings of the 40th International ACM SIGIR Conference on Research and Development in Information Retrieval},
  pages = {1029--1032},
  year = {2017},
  doi = {10.1145/3077136.3080709},
  url = {https://doi.org/10.1145/3077136.3080709}
}

@inproceedings{xiang2025hypermin,
  author = {Xiang, Bin and Cautis, Bogdan and Xiao, Xiaokui and Lakshmanan, Laks V. S.},
  title = {Hyperparametric Influence Minimization: Feature-Driven Intervention Beyond Blocking},
  booktitle = {Proceedings of the 31st ACM SIGKDD Conference on Knowledge Discovery and Data Mining V.2},
  pages = {3297--3308},
  year = {2025},
  doi = {10.1145/3711896.3736999},
  url = {https://doi.org/10.1145/3711896.3736999}
}

@inproceedings{chang2025timin,
  author = {Chang, Xueqin and Fu, Jiajie and Liu, Qing and Gao, Yunjun and Zheng, Baihua},
  title = {Time-Aware Influence Minimization via Blocking Social Networks},
  booktitle = {2025 IEEE 41st International Conference on Data Engineering (ICDE)},
  pages = {557--570},
  year = {2025},
  doi = {10.1109/icde65448.2025.00048},
  url = {https://doi.org/10.1109/icde65448.2025.00048}
}

@inproceedings{shi2019adaptive,
  author = {Shi, Qihao and Wang, Can and Ye, Deshi and Chen, Jiawei and Feng, Yan and Chen, Chun},
  title = {Adaptive Influence Blocking: Minimizing the Negative Spread by Observation-Based Policies},
  booktitle = {2019 IEEE 35th International Conference on Data Engineering (ICDE)},
  pages = {1502--1513},
  year = {2019},
  doi = {10.1109/icde.2019.00135},
  url = {https://doi.org/10.1109/icde.2019.00135}
}

@inproceedings{zhao2019tracking,
  author = {Zhao, Junzhou and Shang, Shuo and Wang, Pinghui and Lui, John C.S. and Zhang, Xiangliang},
  title = {Tracking Influential Nodes in Time-Decaying Dynamic Interaction Networks},
  booktitle = {2019 IEEE 35th International Conference on Data Engineering (ICDE)},
  pages = {1106--1117},
  year = {2019},
  doi = {10.1109/icde.2019.00102},
  url = {https://doi.org/10.1109/icde.2019.00102}
}

@inproceedings{huang2019temporal,
  author = {Huang, Shixun and Bao, Zhifeng and Culpepper, J. Shane and Zhang, Bang},
  title = {Finding Temporal Influential Users Over Evolving Social Networks},
  booktitle = {2019 IEEE 35th International Conference on Data Engineering (ICDE)},
  pages = {398--409},
  year = {2019},
  doi = {10.1109/icde.2019.00043},
  url = {https://doi.org/10.1109/icde.2019.00043}
}

@inproceedings{li2019multifaceted,
  author = {Li, Yuchen and Fan, Ju and Ovchinnikov, George V. and Karras, Panagiotis},
  title = {Maximizing Multifaceted Network Influence},
  booktitle = {2019 IEEE 35th International Conference on Data Engineering (ICDE)},
  pages = {446--457},
  year = {2019},
  doi = {10.1109/icde.2019.00047},
  url = {https://doi.org/10.1109/icde.2019.00047}
}

@inproceedings{khan2016revenue,
  author = {Khan, Arijit and Zehnder, Benjamin and Kossmann, Donald},
  title = {Revenue maximization by viral marketing: A social network host's perspective},
  booktitle = {2016 IEEE 32nd International Conference on Data Engineering (ICDE)},
  pages = {37--48},
  year = {2016},
  doi = {10.1109/icde.2016.7498227},
  url = {https://doi.org/10.1109/icde.2016.7498227}
}

@inproceedings{huang2020profit,
  author = {Huang, Keke and Tang, Jing and Xiao, Xiaokui and Sun, Aixin and Lim, Andrew},
  title = {Efficient Approximation Algorithms for Adaptive Target Profit Maximization},
  booktitle = {2020 IEEE 36th International Conference on Data Engineering (ICDE)},
  pages = {649--660},
  year = {2020},
  doi = {10.1109/icde48307.2020.00062},
  url = {https://doi.org/10.1109/icde48307.2020.00062}
}

@article{xie2024countering,
  author = {Xie, Jiadong and Chen, Zehua and Chu, Deming and Zhang, Fan and Lin, Xuemin and Tian, Zhihong},
  title = {{Influence Maximization via Vertex Countering}},
  journal = {Proceedings of the VLDB Endowment},
  volume = {17},
  number = {6},
  pages = {1297--1309},
  year = {2024},
  doi = {10.14778/3648160.3648171},
  url = {https://doi.org/10.14778/3648160.3648171}
}

@article{huang2024matroid,
  author = {Huang, Yiqian and Zhang, Shiqi and Lakshmanan, Laks V. S. and Lin, Wenqing and Xiao, Xiaokui and Tang, Bo},
  title = {{Efficient and Effective Algorithms for A Family of Influence Maximization Problems with A Matroid Constraint}},
  journal = {Proceedings of the VLDB Endowment},
  volume = {18},
  number = {2},
  pages = {117--129},
  year = {2024},
  doi = {10.14778/3705829.3705833},
  url = {https://doi.org/10.14778/3705829.3705833}
}

@inproceedings{borgs2014rr,
  author = {Borgs, Christian and Brautbar, Michael and Chayes, Jennifer and Lucier, Brendan},
  title = {{Maximizing Social Influence in Nearly Optimal Time}},
  booktitle = {Proceedings of the Twenty-Fifth Annual ACM-SIAM Symposium on Discrete Algorithms},
  pages = {946--957},
  year = {2014},
  doi = {10.1137/1.9781611973402.70},
  url = {https://doi.org/10.1137/1.9781611973402.70}
}

@inproceedings{goyal2010probabilities,
  author = {Goyal, Amit and Bonchi, Francesco and Lakshmanan, Laks V. S.},
  title = {{Learning Influence Probabilities in Social Networks}},
  booktitle = {Proceedings of the Third ACM International Conference on Web Search and Data Mining},
  pages = {241--250},
  year = {2010},
  doi = {10.1145/1718487.1718518},
  url = {https://doi.org/10.1145/1718487.1718518}
}

@article{jin2011distance,
  author = {Jin, Ruoming and Liu, Lin and Ding, Bolin and Wang, Haixun},
  title = {{Distance-Constraint Reachability Computation in Uncertain Graphs}},
  journal = {Proceedings of the VLDB Endowment},
  volume = {4},
  number = {9},
  pages = {551--562},
  year = {2011},
  doi = {10.14778/2002938.2002941},
  url = {https://doi.org/10.14778/2002938.2002941}
}

@article{ke2019reliability,
  author = {Ke, Xiangyu and Khan, Arijit and Quan, Leroy Lim Hong},
  title = {{An In-Depth Comparison of s-t Reliability Algorithms over Uncertain Graphs}},
  journal = {Proceedings of the VLDB Endowment},
  volume = {12},
  number = {8},
  pages = {864--876},
  year = {2019},
  doi = {10.14778/3324301.3324304},
  url = {https://doi.org/10.14778/3324301.3324304}
}

@inproceedings{teng2023iminimize,
  author = {Teng, Siyi and Xie, Jiadong and Zhang, Mingkai and Wang, Kai and Zhang, Fan},
  title = {{IMinimize: A System for Negative Influence Minimization via Vertex Blocking}},
  booktitle = {Proceedings of the 32nd ACM International Conference on Information and Knowledge Management},
  pages = {5101--5105},
  year = {2023},
  doi = {10.1145/3583780.3614743},
  url = {https://doi.org/10.1145/3583780.3614743}
}

@article{mccabe2024deplatforming,
  author = {McCabe, Stefan D. and Ferrari, Diogo and Green, Jon and Lazer, David M. J. and Esterling, Kevin M.},
  title = {{Post-January 6th deplatforming reduced the reach of misinformation on Twitter}},
  journal = {Nature},
  volume = {630},
  number = {8015},
  pages = {132--140},
  year = {2024},
  doi = {10.1038/s41586-024-07524-8},
  url = {https://doi.org/10.1038/s41586-024-07524-8}
}
\endgroup

\end{document}